\documentclass{myarticle}
\usepackage{float}
\usepackage{booktabs}
\usepackage{threeparttable}

\usepackage{amsmath}
\usepackage{lineno}
\usepackage{mathtools}
\usepackage{supertabular}
\usepackage{graphicx}
\usepackage{mathrsfs}
\usepackage{amssymb}
\usepackage{caption}
\usepackage{multirow}
\usepackage{array}
\usepackage{url}
\usepackage{color}
\usepackage{bbm}
\usepackage[perpage,symbol]{footmisc}
\usepackage[position=t,singlelinecheck=off]{subfig}

\newcommand{\beginsupplement}{
        \setcounter{table}{0}
        \renewcommand{\tablename}{\textbf{Supplementary Table}}
        \renewcommand{\thetable}{\textbf{\arabic{table}}}
        \setcounter{figure}{0}
        \renewcommand{\figurename}{\textbf{Supplementary Figure}}%
        \renewcommand{\thefigure}{\textbf{\arabic{figure}}}
}
\setfnsymbol{wiley}
            
\title{Discrimination universally determines reconstruction of multiplex networks}

\author{Mincheng Wu$^{1}$, Jiming Chen$^1$, Shibo He$^1$, Youxian Sun$^1$, Shlomo Havlin$^2$, Jianxi Gao$^{3,4}$}
\begin{document}

\maketitle

\begin{affiliations}
\item State Key Laboratory of Industrial Control Technology, Zhejiang University, Hangzhou 310027, China.
\item Department of Physics, Bar-Ilan University, Ramat-Gan 52900, Israel.
\item Department of Computer Science, Rensselaer Polytechnic Institute, Troy, NY 12180, USA.
\item Network Science and Technology Center, Rensselaer Polytechnic Institute, Troy, NY 12180, USA.
\end{affiliations}

\begin{center}
(\date{\today})
\end{center}

\begin{center}
\textbf{Abstract}
\end{center}
\begin{abstract}
Network reconstruction is fundamental to understanding the dynamical behaviors of the networked systems.
Many systems, modeled by multiplex networks with various types of interactions, display an entirely different dynamical behavior compared to the corresponding aggregated network.
In many cases, unfortunately, only the aggregated topology and partial observations of the network layers are available, raising an urgent demand for reconstructing multiplex networks.
We fill this gap by developing a mathematical and computational tool based on the Expectation-Maximization framework to reconstruct multiplex layer structures. The reconstruction accuracy depends on the various factors, such as partial observation and network characteristics, limiting our ability to predict and allocate observations. Surprisingly, by using a mean-field approximation, we discovered that a discrimination indicator that integrates all these factors universally determines the accuracy of reconstruction. This discovery enables us to design the optimal strategies to allocate the fixed budget for deriving the partial observations, promoting the optimal reconstruction of multiplex networks. To further evaluate the performance of our method, we predict beside structure also dynamical behaviors on the multiplex networks, including percolation, random walk, and spreading processes. Finally, applying our method on empirical multiplex networks drawn from biological, transportation, and social domains, corroborate the theoretical analysis.

\end{abstract}

\clearpage
Multiplex networks, composed of a collection of layers sharing the same node-set, can describe multiple types of interactions between nodes in different layers more precisely than the corresponding aggregated networks~\cite{boccaletti2014structure,kivela2014multilayer,bianconi2018multilayer,moreno2019focus,cozzo2018multiplex}.
Neglecting the multiplex structure may lead to inaccurate consequences since dynamical behaviors on a multiplex network significantly differ from those on its aggregated one~\cite{de2015ranking,zeng2019switch,cardillo2013emergence,wu2019tensor}.
For example, a tiny fraction of node removal in one layer may cause cascading failures between layers and the catastrophic collapse of the entire multiplex network~\cite{buldyrev2010catastrophic} (Fig. 1A). 
However, the corresponding aggregated network may remain unscathed for the same set of node removal (Fig. 1B). 
The recent results corroborate that an interdependent network is more vulnerable than its aggregated against random failures~\cite{gao2011robustness,danziger2019dynamic}.
Further, the cost of switching between two layers determines the navigability of a multiplex transportation network~\cite{de2014navigability}. 
Another example is that a random walk process over a multiplex transportation network, the walker may need an extra cost to switch from one layer to other layers (Fig. 1C). In contrast, one can walk along with any link without additional cost within the aggregated network (Fig. 1D). 
Such an extra cost causes a lower coverage of the nodes that are visited by the walker in a multiplex transportation network. 
A further example is the dynamic of the spreading process in a temporal network, modeling rumor circulation in a social network or flu outbreak in a susceptible population network~\cite{pastor2015epidemic,li2019impacts,alvarez2019dynamic}.
The topology of a temporal network may change at each time (Fig. 1E). However, it is static in the aggregated network (Fig. 1F), which will lead to a lower infected fraction in a temporal network than that in the aggregate one.

Unfortunately, it is practically difficult to obtain data representing accurately the layers of the multiplex topology. 
Instead, the only aggregate topology of multiple layers and partial observations (e.g., a sampled subnetwork) in each layer are available. 
It is costly, time-consuming, and in many cases impossible to measure all types of interactions and heterogeneity of nodes, especially in a large-scale complex system. 
Researchers, for example, can construct the entire connectome of Caenorhabditis Elegans' neural system~\cite{yan2017network,cook2019whole} (Fig. S1), potentially offering a better understanding of brains' functionality. 
This potential is limited because of the unidentified multiplex topology, reflecting on the types of interactions (e.g., gap-junction or synapse) between any two connected neurons without immense experiments~\cite{white1986structure}.
Analogous cases widely appear in various aspects of life, including social networks~\cite{gauvin2013activity} and transportation networks~\cite{lacasa2018multiplex}, pressingly promoting new tools that can leverage limited prior knowledge for accurately and efficiently reconstructing multiplex network.

The ultimate goal of network reconstruction is to estimate the topology of the system using limited observations, which reveals the dynamical behaviors of the original system. Towards this goal, much efforts has been devoted to find the most probable microscale structure or mesoscale structure based on incomplete date~\cite{peel2017ground,newman2016structure,massucci2016inferring,taylor2017super}. Examples are predicting potential links that may appear in evolving networks~\cite{lu2011link}, inferring link reliability~\cite{guimera2009missing}, recovering missing links or locating spurious links~\cite{newman2018networkstr,newman2018networkrec}.
Recent works have met with success in detecting if an observed single-layer network is an aggregation of a hidden multiplex structure~\cite{valles2016multilayer} or if an observed dynamical process (e.g., a random walk process) is an aggregation of several processes taking place on different hidden layers~\cite{lacasa2018multiplex}.
Since many real-world networks have multiplex structures, it is also crucial to reconstruct multiplex networks, which can predict links that may appear in the future by the current multiplex structure for example~\cite{de2017community}.
Nevertheless, as far as we know, a framework to reconstruct multiplex layer structures, displaying the specific structure of each hidden layer is still lacking.

Several notable challenges prevent us from addressing multiplex network reconstruction.
First of all, the same aggregate network can be generated by different combinations of single-layer networks. There exists an enormous number of possible mappings from the potential multiplex layers structures to the observed aggregate topology.
Specifically, the probability space composed by potential multiplex structures has an exponential ($(2^L-1)^{|A^\mathcal{O}|}$) possibilities with the number of layers $L$ and the number of observed links $|A^\mathcal{O}|$ (see Fig. S2 for more details).
Can one conceive a low-complexity framework to reconstruct multiplex layer structures, avoiding the enormous cost of ergodic methods~\cite{santoro2019algorithmic}?
It is practically infeasible to know the generating models of the multiplex networks or the dynamical processes on them. 
It, therefore, fails if we reconstruct networks along with the state of the art methods based on specific models or dynamic processes.
How to develop a universal framework to reconstruct the multiplex layer structures and further reveal the original dynamical behaviors solely by knowing the available aggregate topology as well as limited partial observations of layers, removing the long-standing constraints? Moreover, various characteristics of the multiplex structure affect the performance of reconstruction. For example, the disparity of average degrees and overlap of edges in different layers have a different impact on the performance of reconstruction. 
Is there an indicator universally quantifying the fundamental relation between the reconstruction accuracy and diverse network characteristics?
Lastly, more partial observations yield higher reconstruction accuracy while they incur a more cost.
However, there are too many possibilities to allocate a limited budget in different layers.
Is there an optimal strategy to allocate the limited budget that enables the highest accuracy of reconstruction for various multiplex networks by the indicator? 
To answer these fundamental questions, in this article, we propose a mathematical and computational framework that can reconstruct multiplex network and predict the dynamic process on it. We found a discrimination indicator based on information entropy, integrated by multiple network characteristics, that universally determines the reconstruction accuracy of multiplex networks. This discovery enables us to design the optimal strategy to allocate a fixed budget for partial observations, promoting the optimal reconstruction of multiplex networks. Experimental results based on nine real-world multiplex networks and several synthetic networks corroborate our analytical results.

\clearpage
\section*{Results}
\subsection*{Framework for reconstructing multiplex layer structures}
Suppose that we have an aggregate topology $A^\mathcal{O}$ and partial observations $\Gamma$ (i.e., a subgraph in each layer) from a multiplex network $\mathbf{M}$ that is composed of $L$ layers and $N$ nodes in each layer (Fig. 2A).
We denote the percentage of partial observations by $c$ ($0\leq c<1$), indicating the proportion of edges in the observed subgraph to those in the whole network.
We clarify the framework with multiplex networks aggregated by the OR mechanism, which is the most common case ranging from biological networks to social networks (see Supplementary Text A for more aggregate mechanisms).
We employ the configuration model to measure the probability that an edge exists between any two nodes, which exploits an arbitrary degree sequence $\vec{d}\in \mathbbm{R}^N$ to describe a network.
The configuration model can significantly reduce the complexity from exponential to polynomial by exploiting the independence of each link and has been widely applied to analyze the relationship between structure and function of complex networks~\cite{newman2003structure,courtney2016generalized,newman2018networks}.

The first step is to find the most probable values of $\vec{\mathbf{d}}$ by maximizing the posterior probability $P(\vec{\mathbf{d}}|A^\mathcal{O},\Gamma)$, where $\vec{\mathbf{d}}=(\vec{d^1},\vec{d^2},\cdots,\vec{d}^L)$ encodes the expected degree sequences in all layers.
Since there is no prior knowledge about the degrees $\vec{\mathbf{d}}$, we assume a uniform distribution for the prior, i.e., $P(\vec{\mathbf{d}})=constant$~\cite{peixoto2013parsimonious}. Note that this method can be improved if we know some prior knowledge about the degrees.
Based on the Bayesian rule, the maximum posterior estimate is equivalent to maximizing the likelihood function $l(\vec{\mathbf{d}})=P(A^\mathcal{O},\Gamma|\vec{\mathbf{d}})$, which performs the maximum likelihood estimation (MLE). Employing the law of total probability, we have
\begin{equation}
l(\vec{\mathbf{d}})=\sum_{\mathbf{M}}P(A^\mathcal{O},\Gamma|\mathbf{M})\cdot P(\mathbf{M}|\vec{\mathbf{d}}),
\label{eq.Pd}
\end{equation}
which is a summation over all possible potential multiplex structure $\mathbf{M}$.
As various potential multiplex structures produce the same aggregate topology, we denote the probability distribution for all multiplex structure by $Q(\mathbf{M})$, and $\sum_\mathbf{M} Q(\mathbf{M})=1$.
Then, the estimated degrees $\vec{\mathbf{d}}$  can reconstruct the multiplex structure by calculating the posterior distribution
\begin{equation}
\begin{centering}
Q(\mathbf{M})=P(\mathbf{M}|A^\mathcal{O},\Gamma,\vec{\mathbf{d}}).
\end{centering}
\label{eq.QM}
\end{equation}
Note that $P(A^\mathcal{O},\Gamma|\vec{\mathbf{d}})$ and $Q(\mathbf{M})$ are interdependent, and thus we perform an iterative process to obtain the MLE of the degrees $\vec{\mathbf{d}}$ and the posterior distribution $Q(\mathbf{M})$ as follows.
Given a guessed initial value $\vec{\mathbf{d}}^{(0)}$, we find the optimized posterior distribution $Q^{(k)}(\mathbf{M})$ in Eq. (\ref{eq.QM}) by $\vec{\mathbf{d}}^{(k-1)}$. 
Then, we update the parameters $\vec{\mathbf{d}}^{(k)}$ that maximize the Eq. (\ref{eq.Pd}) by posterior distribution $Q^{(k)}(\mathbf{M})$, which perform a coordinate ascent to maximize the likelihood (Fig. 2B) (see Supplementary Text B for the complete algorithm). 
The iterations above are derived from the expectation-maximization (EM) algorithm~\cite{dempster1977maximum} (see Methods and Materials), and a toy example is shown (Fig. 2C).
Notice that the likelihood function Eq. (\ref{eq.Pd}) will be replaced by the product of the likelihood function $P(A^\mathcal{O},\Gamma|\vec{\mathbf{d}})$ and the prior $P(\vec{\mathbf{d}})$ if there is any prior on the parameters $\vec{\mathbf{d}}$, which performs the  maximum a prior estimation (MAP) then.

In estimation and statistics theory, an unbiased estimator is called efficient if the variance of the estimator reaches Cramer-Rao lower bound (CRLB)~\cite{lehmann2004elements}.
Fortunately, the proposed framework yields a maximum likelihood estimation, which is an unbiased estimator, and performs asymptotic normality indicating the estimator converges in distribution to a normal distribution~\cite{newey1994large}. 
With this, we prove that the variance of the estimator designed in our framework decreases as the percentage of partial observations $c$ increases, and further reaches the CRLB when the network size $N$ approaches infinity (see Supplementary Text C and Fig. S3 for more details).

\subsection*{Evaluations for performance of reconstruction}
We now analyze the performance of reconstruction on various real-world multiplex networks. 
After estimating degree sequences $\vec{\mathbf{d}}$, the posterior probability $Q^{\alpha}_{ij}$ can be calculated by
\begin{equation}
Q^{\alpha}_{ij}=P(M^{\alpha}_{ij}=1|A^\mathcal{O},\Gamma,\vec{\mathbf{d}}),
\end{equation}
which is called link reliability, indicating the probability that a link exists between node $i$ and node $j$ in layer $\alpha$.  
We examine the reliability of all links in testing set $E^T$ consisting of potential links except partial observations, i.e., $E^T=\{M^\alpha_{ij}\notin \Gamma|A^{\mathcal{O}}_{ij}=1\}$ (see Fig. S4 for more details).
For this purpose, we calculate the TP (true positive rate) $P(M_{ij}^{\alpha}=1|Q^{\alpha}_{ij}>q)$, FP (false positive rate) $P(M_{ij}^{\alpha}=0|Q^{\alpha}_{ij}>q)$, TN (true negative rate) $P(M_{ij}^{\alpha}=0|Q^{\alpha}_{ij}<q)$ and FN (false negative rate) $P(M_{ij}^{\alpha}=1|Q^{\alpha}_{ij}<q)$ in $E^T$, where $q$ is an application-dependent threshold.

We first set the threshold $q=0.5$, and calculate the four metrics to evaluate the performance of multiplex reconstruction for the nine real-world datasets (see table S1 for more details of the real-world datasets).
The percentage of partial observations $c$, indicating the portion of the observed edges, exhibits a positive correlation with the accuracy of reconstruction, showing good performance even with a quite small $c$ (Fig. 3A) (see Methods and Materials and Fig. S5 for more evaluations). 
Then, we range the threshold $q$ from 0 to 1, which determines the classifier boundary for varying classes, and display the ROC curves against different $c$ in the ROC space (Fig. 3, B and C). 
Here, the ROC space is defined by false positive rate and the true positive rate as horizontal and vertical axes, respectively, displaying the relative trade-offs between false positive (costs) and true positive (benefits). 
Further, the true positive rate is positively correlated with false positive rate, and there exists a threshold, above which a false positive rate increases faster than the true positive rate. 
It is, thereby, not judicious anymore to improve a true positive rate by increasing the false positive rate beyond such a threshold. 
 
Network properties also include average degree, degree distribution, length of the shortest path, which are significant to network reconstruction. 
One prominent advantage of the reconstruction framework is that we can simultaneously obtain other micro- or mesoscale network properties.
For example, the expectation of the degree distribution of layer $\alpha$ is obtained by $\mathbf{E}(p_\alpha)=\sum_\mathbf{M}Q(\mathbf{M})\cdot p_\alpha(\mathbf{M})$, where $p_\alpha(\mathbf{M})$ is the degree distribution of layer $\alpha$ in multiplex network $\mathbf{M}$.
The degree distributions in two layers for different $c$ are compared to the real multiplex network, demonstrating that the degree distributions in all layers are well reconstructed as $c$ increases (Fig. 3, D and E).
Generally, the expectation of property $X$ is the first raw moment obtained by $\mathbf{E}(X)=\sum_\mathbf{M}Q(\mathbf{M})\cdot X(\mathbf{M})$, while the corresponding variance is the second central moment $\mathbf{D}(X)=\sum_\mathbf{M}Q(\mathbf{M})\cdot \left[X(\mathbf{M})-\mathbf{E}(X)\right]^2$. 
Moreover, we can obtain skewness, kurtosis, and higher moments of a property $X$.
We will discuss how different network characteristics (e.g., average degrees and overlap of edges in different layers) impact the performance of reconstruction in the next section.

\subsection*{The universal discrimination indicator}
It is also interesting to investigate how various characteristics of multiplex networks affect the performance of reconstruction.
Without loss of generality, we conduct an analysis of two-layer multiplex networks for clarification.
The probability space of potential multiplex structure contains three events: the link exists (i) only in layer 1, (ii) only in layer 2, or (iii) in both layers when a link between node $i$ and node $j$ is observed in the aggregate network.
Then, the uncertainty of all links in the multiplex network can be quantified by the entropy
\begin{equation}
\mathcal{H}(\mathbf{M}|A^\mathcal{O})=\mathcal{H}(M^1|A^\mathcal{O})+\mathcal{H}(M^2|A^\mathcal{O}),
\end{equation}
where $M^1$ and $M^2$ are the adjacency matrices of the two layers in a multiplex network.
Generally speaking, the smaller the entropy $\mathcal{H}$ is, the more certain the potential multiplex structure is, and vice versa.

To study how different characteristics of multiplex networks impact on the entropy, we first introduce the ratio of average degrees of two layers denoted by $r$, i.e., $ r=\frac{\left \langle k_1 \right \rangle}{\left \langle k_2 \right \rangle }$, where $\left \langle k_1 \right \rangle$ and $\left \langle k_2 \right \rangle$ are the average degrees of layer 1 and layer 2, respectively.
We assume $\left \langle k_1 \right \rangle\leq\left \langle k_2 \right \rangle$ without loss of generality, such that $0< r\leq 1$.
Then, we consider the overlap of edges denoted by $v$ between the two layers.
A high overlap indicates that a link is more likely to exist in one layer if the corresponding link exists in the other layer, i.e., a low uncertainty. 
To measure the overlap $v$ of a multiplex network, we refer to the Jaccard index of $E_1$ and $E_2$ indicating the two edge sets in the two layers, i.e., $v={|E_1 \cap E_2|}/{|E_1\cup E_2|}$ (see Fig. S7 for more details about multiplex network characteristics).

We next explore how these factors impact the entropy of a multiplex network and further determine the performance of reconstruction.
We can calculate the entropy $\mathcal{H}$ by the mean-field approximation (see Methods and Materials for more details), and obtain
\begin{equation}
\mathcal{H}(\mathbf{M}|A^\mathcal{O})= -\frac{N(N-1)}{2}\cdot\sum^2_{\alpha=1}[p_\alpha\cdot\ln p_\alpha+(1-p_\alpha)\cdot\ln (1-p_\alpha)],
\label{eq.H}
\end{equation}
where 
\begin{equation}
p_\alpha=\left\{
             \begin{array}{lr}
             \frac{\hat v+\hat r}{1+\hat r}, \ \text{if}\ \alpha=1\\
             \frac{1+\hat v\cdot\hat r}{1+\hat r},\ \text{if}\ \alpha=2
             \end{array}
\right..\label{eq.palpha}
\end{equation}
Notice that $\hat v$ and $\hat r$ are the estimations when we only have partial observations $\Gamma$, and we approximate them by $c\cdot v$ and $r^c$ empirically.
Thus, we find that the entropy of a given multiplex network is highly related to the percentage of partial observations $c$, the ratio of average degrees $r$ and overlap $v$. 
It is clear that the uncertainty of the probability space decreases with $c$ and $v$ increasing.
Hence, the entropy $\mathcal{H}$ is a monotonously decreasing function of $c$ and $v$ over the domain.
For $r$, however, the entropy is a monotonously increasing function when $r$ increases from 0 to 1 (Fig. 4A).
Clearly, $\mathcal{H}$ describes the microscale discrimination between layers of a multiplex network, since a high discrimination ($r$ tends to 0) indicates a low entropy.

Generally, the accuracy is expected to be determined by entropy $\mathcal{H}$, since the entropy $\mathcal{H}$ is the primary variable that determines the uncertainty of the potential space of multiplex structure. 
Empirically, we find that the accuracy of reconstruction is universally in direct proportion to the indicator $1-\rho\cdot\mathcal{H}$ (Fig. 4B), i.e.,
\begin{equation}
\text{Accuracy} \propto  1-\rho\cdot\mathcal{H},
\label{eq.ACC}
\end{equation}
where $\rho$ is a scaling factor satisfying 
\begin{equation}
\rho=\frac{1}{2\ln2\cdot N(N-1)}\cdot(1-\frac{1-v}{1+v}\cdot c^s).
\label{eq.rho}
\end{equation}
In Eq. (\ref{eq.rho}), $s=s_{(\mathbf{M})}$ is a constant related to the given topology of the multiplex network $\mathbf M$ (see table S2 for approximate values of $s_{(\mathbf{M})}$.
The term ${(1-v)}/{(1+v)}$ in the Eq. (\ref{eq.rho}) indicates the uncertainty of links in testing set can be reduced by partial observations, and $s$ describes the scale of partial observations can reduce the uncertainty of links in testing set (see Methods and Materials).
We further find that $s_{(\mathbf{M})}$ is highly proportional to the cosine similarity of the two degree sequences in each layer, i.e., $s \propto\cos \left \langle \vec{d^1},\vec{d^2} \right \rangle$ (Fig. 4C). 
Clearly, $\cos \left \langle \vec{d^1},\vec{d^2} \right \rangle$ describes the similarity between degree sequences of two layers in a multiplex network, indicating the mesoscale discrimination, which is not relevant to microscale discrimination including $r$, $v$, and $\mathcal{H}$ generally.

Thus, the accuracy of reconstruction is determined by the universal discrimination indicator ($1-\rho\cdot \mathcal{H}$) from both microscale and mesoscale views.
This discovery indicates that the reconstruction can be predicted accurately by the discrimination indicator, obtaining a high accuracy of reconstruction either $\rho$ and $\mathcal{H}$ is small.
For example, the accuracy of reconstruction can be enhanced when the difference in average degrees between layers is vast ($r$ tends to $0$).
Notice that we can approximate $s$ by the cosine similarity if we do not meet the exact value of $s$ empirically, since ${s}$ is highly related to the cosine similarity.
We will next discuss how to allocate the partial observations in different layers when a specific budget $\bar c$ is given.

\subsection*{Allocating limited budget for partial observations}
Usually, we have a limited budget for conducting observations in practice.
It is, thereby, interesting to investigate budget allocation (partial observations $\Gamma$) in different layers to optimize the performance of reconstruction (e.g., the accuracy) as far as possible.
We denote the average partial observations by $\bar{c}$, i.e., $\bar{c}=\sum_\alpha c_\alpha/L$, where $c_\alpha$ indicates the percentage of partial observation in layer $\alpha$, and denote $\mathcal{H}(M^\alpha|A^\mathcal{O})$ by $\mathcal{H}_\alpha$ for simplicity.
Similarly, employing the mean-field approximation (see Methods and Materials), we can predict the accuracy by the function $F$ defined as 
\begin{equation}
F(c_1,c_2) = 1-\frac{1-c_1}{1-c_1+(1-c_2)/ \hat r}\cdot\rho_1\cdot\mathcal{H}_1-\frac{(1-c_2)/ \hat r}{1-c_1+(1-c_2)/ \hat r}\cdot\rho_2\cdot\mathcal{H}_2,
\label{eq.ACC2}
\end{equation}
where
\begin{equation}
\rho_\alpha=\frac{1}{2\ln2\cdot N(N-1)}\cdot(1-\frac{1-v}{1+v}\cdot c_{3-\alpha}^{s}),\ \alpha=1,2.
\label{eq.rho1}
\end{equation}
We next explore how the performance of reconstruction is impacted by different ratio $c_1/c_2$ when given a certain budget for average partial observations.
Once $\bar{c}$ is given, we regard the function $F$ as a unary function of $c_1$, i.e., $F=F(c_1)$, since $c_2=2\bar c-c_1$.
Then, the domain of $F(c_1)$ is $[0,2\bar c]$ if $\bar c\leq 0.5$, and is $[2\bar c-1,1]$ if $\bar c> 0.5$.

We notice that the function $F(c_1)$ monotonically increases over the domain if $\bar c $ is small, but decreases at first and increases late if $\bar c $ is large.
Theoretical analysis shows that $F(0)\geq F(2\bar c)$ if $\bar c\leq 0.5$, and $F(2\bar c-1)\geq F(1)$ if $\bar c> 0.5$ (see Methods and Materials).
The result indicates that it is always better to allocate the budget as much as possible to the layer whose average degree is lower, and  we can reach the optimal strategy to obtain the highest accuracy then.
Moreover, there exists a threshold $0<\bar{c}_0(\mathbf{M})<1$ for each multiplex network $\mathbf{M}$, where $\bar c_0$ is the solution to the equation
\begin{equation}
F(0)=F(2\bar c_0).
\label{eq.c_0}
\end{equation}
If the budget $\bar{c}$ is less than $\bar c_0$, the accuracy increases when $c_1/c_2$ increases, and reaches the maximum as $c_1/c_2$ tends to $\infty$.
If the budget $\bar{c}$ is large ($\bar{c}>\bar c_0$), however, the accuracy increases when $c_1/c_2$ tends to 0 or $\infty$, and reaches the maximum as $c_1/c_2$ tends to $\infty$ (Fig. 5A), indicating that the multiplex network can be reconstructed when the aggregate topology and either of the two layers is observed. 
The reason is as follows. 
The partial observations in different layers can capture the maximal characteristics of each layer when $c_1/c_2=1$.
However, it will lead to more redundancy and lower accuracy if the partial observations in different layers have a high overlap of observations, making the performance even worse when $c_1/c_2=1$ and $\bar{c}$ is large.
The theoretical results enable us to make the best strategy to allocate budget and thus obtain the optimal reconstruction of multiplex networks.
Furthermore, results from real-world data sets verified our theoretical analysis (Fig. 5, A and B).
We will discuss how different multiplex network characteristics impact the performance of reconstruction from a dynamical behavior point of view in the next section.

\subsection*{Predicting dynamic processes in multiplex networks}
We proceed to investigate the performance of the reconstructed multiplex networks on the prediction of dynamic processes, which is critical to the network functionality.
First, we study a percolation process occurring on a two-layer interdependent multiplex network.
In such a multiplex network, once a set of nodes is removed (e.g., being attacked or random failure) in one layer, nodes disconnected to the GCC in the same layer and the counterparts of the removed nodes will also fail and thus be removed. 
The new removed nodes result in more node removal, and the repetitive processes lead to the catastrophic cascade of failures.
For the reconstructed multiplex network encoded by the expectation $\mathbf{E}[Q(M)]$, we binarize the matrix $\mathbf{E}[Q(M)]$ and randomly remove nodes in one layer with probability $1-p$ (see Supplementary Text D for more details of the process).
We calculate the size of GMCC as a function of $p$ and the critical threshold $p_c$, above which the GMCC exists. 
We compare the average size of GMCC in the reconstructed network (repeated 100 times) to the real one with the C. Elegans neural network against ranging $c$ (Fig. 6A).
The performance of reconstruction is well as shown from the size of GMCC, even if $c$ is small. 
The estimates of the size of GMCC and the critical probability $p_c$ approach those of the real networks ($c=1$) as $c$ increases 1.
However, the proposed method slightly underestimates both the size of GMCC and the critical threshold $p_c$ for the C. Elegans neural network.
Further, simulations on synthetic networks reveal that the method underestimates much more the robustness and $p_c$ of the interdependent networks when $r$ is small and closes to 0 (Fig. 6B).

Second, we consider a random walk process taking place on interconnected multiplex networks, where interlayer links only exist between counterparts.
We suppose that a number of walkers start from randomly chosen nodes and walk along with intralayer links with a probability $p_{intra}$, and along with interlayer links with probability $p_{inter}$ (see Supplementary Text E for more details).
We employ the coverage $\phi(t)$ as the performance metric, indicating the proportion of nodes that have been visited by the walkers before time $t$. 
The coverage at each time on reconstructed multiplex networks are compared to the real one (London multiplex transportation network) against different $c$, showing an outstanding prediction as $c$ increases (Fig. 6C).
Simulations on synthetic networks show that the multiplex networks will be overestimated no matter $r$ is small or large (Fig. 6D).

Last, we investigate a spreading process based on the SI (susceptible-infected) model on temporal networks, where interlayer links only exist between two counterparts at two adjacent times~\cite{gauvin2013activity}.
In the SI model, each node has only two states: ``susceptible'' (S) or ``infected'' (I), and at initial time $t=0$, $5\%$ nodes are randomly chosen to be sources (infected).
At each time (which corresponds to one layer in the temporal network), the infected nodes will infect the susceptive neighboring nodes with a specific infection rate of $\lambda$ (see Supplementary Text F for more details).
In the spreading process, the proportion of infected nodes $I(T)$ at time $t=T$ on reconstructed networks is calculated and compared to the real one with the social interactions at SFHH (La Société française d'Hygiène Hospitalière) conference (Fig. 6E). 
Interestingly, the process taking place on multiplex networks with a small $r$ will be predicted well (Fig. 6F).
Moreover, we have studied how the performance of reconstruction for dynamics is influenced by more network characteristics, including the overlap of edges and ratio of heterogeneity (see Fig. S8 for more results on real-world networks and Fig. S9 for more results on synthetic network with different characteristics).

\section*{Discussion}
Network reconstruction has attracted much research attention recently and has wide applications such as link prediction, community detection and systems' vulnerability analysis. 
Most previous studies focused on  monoplex networks, and therefore there is a pressing need to develop a reconstruction framework for multiplex networks. 
Existing work have met success to determine if an observed monoplex network is the outcome of a hidden multilayer process by assuming generative models for each layer of the multiplex network.
However, it is necessary to further explore the multiplex structure and predict the dynamics if it is verified that there is a hidden multiplex structure.
Our primary goal is to reconstruct the multiplex structure from the knowledge of an observed aggregate monoplex network and partial observations.

However, there are many challenges preventing us to build a framework for reconstruction of multiplex networks. 
Given the aggregation mechanism (e.g., the OR mechanism), apparently, there are a large number of potential structures given the aggregate network. 
To avoid the ergodic methods, we propose a framework by building a probability space $Q(\mathbf{M})$, and reduce the complexity from exponential to polynomial by employing the configuration model that allows an arbitrary degree sequence.
Since priors on generating models or other dynamic information are almost unavailable, while the local subgraph information (referred to as partial observations in this article) in some specific layers is more accessible. 
For this purpose, we need to estimate the node degree sequence $\vec{\bf d}$ based on very limited partial observations and, unfortunately it is interdependent on the posterior probability distribution $Q(\bf M)$. 
We design an efficient mathematical framework based on Expectation-Maximization method performing a maximum likelihood estimation, and  prove that the variance of the estimation reaches the CRLB when network size $N$ approaches infinity. 
We evaluate the performance of the reconstruction using various empirical multiplex networks, ranging from microscale (e.g., accuracy of link reliability) to mesoscale (e.g., degree distributions).
Experimental results demonstrate that the performance of reconstruction mounts quickly initially with a small percentage of partial observations, exhibiting the power of the proposed reconstruction framework. 
By the mean-field approximation, surprisingly, we find that a discrimination indicator that integrates all considerable factors universally determines the accuracy of reconstruction, which theoretically aids us to have a deep understanding between the accuracy and the network characteristics.
Thus, the indicator enable us to make the best strategy to allocate limited budget and further obtain the highest accuracy, i.e., the optimal reconstruction.
We also investigate the performance from dynamical view, finding that the  proposed framework can well predict dynamic processes taking place on multiplex networks, and the impact of network characteristics (e.g., average degree, heterogeneity and overlap of edges in different layers) on performance is analyzed.

To the best of our knowledge, we provide the most comprehensive mathematical framework for reconstructing the multiplex network layer structures. 
It paves a new way of understanding the structure and function in multiplex networks, and our discovery reveals the essential feature of multiplex network reconstruction.
Our future work will focus on unveiling other orthogonal knowledge to further improve the reconstruction performance.
We believe that the proposed framework will have a broader impact in many different applications including link prediction,  missing links recovery,  spurious links location drawn from biological, social, transportation domains.

\clearpage
\section*{Methods and Materials}
\subsection*{Expectation maximization framework}

In this section we will present details on how to obtain the maximum likelihood estimation of $\vec{\mathbf{d}}$.
We rewrite the likelihood function
\begin{equation}
P(A^\mathcal{O},\Gamma|\vec{\mathbf{d}})=\sum_{\mathbf{M}}P(A^\mathcal{O},\Gamma|\mathbf{M},\vec{\mathbf{d}})P(\mathbf{M}|\vec{\mathbf{d}}),
\end{equation}
by employing the law of total probability (summing over all possible multiplex structure $\mathbf{M}$).
The primary objective is to find $\vec{\mathbf{d}}$ that maximizes the likelihood above.
In practice, we will maximize its logarithm $\ln P(A^\mathcal{O}, \Gamma|\vec{\mathbf{d}})$ rather than $P(A^\mathcal{O}, \Gamma|\vec{\mathbf{d}})$ for the purpose of convenience. 
Clearly, 
\begin{equation}
\ln{P(A^\mathcal{O},\Gamma|\vec{\mathbf{d}}})=\ln\sum_{\mathbf{M}}P(A^\mathcal{O},\Gamma,\mathbf{M}|\vec{\mathbf{d}}).
\end{equation}
Employing the Jensen's inequality, we have
\begin{equation}
\ln\sum_{\mathbf{M}}P(A^\mathcal{O},\Gamma,\mathbf{M}|\vec{\mathbf{d}})\geq\sum_{\mathbf{M}}Q(\mathbf{M})\ln\frac{P(A^\mathcal{O},\Gamma,\mathbf{M}|\vec{\mathbf{d}})}{Q(\mathbf{M})},
\label{eq.Ji}
\end{equation}
where $Q(\mathbf{M})$ is an arbitrary distribution of the multiplex structure $\mathbf{M}$ satisfying $\sum_{\mathbf{M}}Q(\mathbf{M})=1$.
For simplicity, we denote 
\begin{align}
&J(Q,\vec{\mathbf{d}})=\sum_{\mathbf{M}}Q(\mathbf{M})\ln\frac{P(A^\mathcal{O},\Gamma,\mathbf{M}|\vec{\mathbf{d}})}{Q(\mathbf{M})},
\label{eq.J}
\end{align}
which is a lower bounding function of $\ln P(A^\mathcal{O}, \Gamma|\vec{\mathbf{d}})$. Notice that $J$ is a function of the distribution $Q(\mathbf{M})$ and the parameters $\vec{\mathbf{d}}$.

In the expectation-maximization (EM) algorithm~\cite{dempster1977maximum}, we will maximize the function $J$  by recursively executing two steps: E-step and M-step. 
In the E-step, we maximize $J(Q,\vec{\mathbf{d}})$ while keeping $\vec{\mathbf{d}}$ as constants. It is easy to see that  Eq. (\ref{eq.Ji}) holds  if and only if $Q(\mathbf{M})$ is the posterior distribution of the multiplex structure $\mathbf{M}$, i.e.,
\begin{align}
Q(\mathbf{M})&=\frac{P(A^\mathcal{O},\Gamma,\mathbf{M}|\vec{\mathbf{d}})}{\sum\limits_{\mathbf{M}}P(A^\mathcal{O},\Gamma,\mathbf{M}|\vec{\mathbf{d}})}=\frac{P(A^\mathcal{O},\Gamma,\mathbf{M}|\vec{\mathbf{d}})}{P(A^\mathcal{O},\Gamma|\vec{\mathbf{d}})}=P(\mathbf{M}|A^\mathcal{O},\Gamma,\vec{\mathbf{d}}).
\end{align}
In the M-step, we differentiate Eq. (\ref{eq.J}) with respect to $\vec{\mathbf{d}}$ while fixing $Q{(\mathbf{M})}$, and find the solution to the following equation
\begin{equation}
\frac{\partial}{\partial\vec{\mathbf{d}}} \sum_{\mathbf{M}}Q(\mathbf{M})\ln P(A^\mathcal{O},\Gamma,\mathbf{M}|\vec{\mathbf{d}})=0.
\end{equation}
Notice that $\sum_{\mathbf{M}}Q(\mathbf{M})\ln P(A^\mathcal{O},\Gamma,\mathbf{M}|\vec{\mathbf{d}})$ is the posterior expectation of the logarithmic likelihood function $\ln P(A^\mathcal{O},\Gamma,\mathbf{M}|\vec{\mathbf{d}})$ with respect to the distribution $Q{(\mathbf{M})}$.
Thus, given guessed initial parameters, we iteratively update the distribution $Q(\mathbf{M})$ and parameters $\vec{\mathbf{d}}$  until they converge. 
The two steps can be written as the iteration scheme
\begin{equation}
\begin{cases}
Q^{(k)}(\mathbf{M})&=P(\mathbf{M}|A^\mathcal{O},\Gamma,\vec{\mathbf{d}}^{(k)}).\\
\vec{\mathbf{d}}^{(k+1)}&=\underset{\vec{\mathbf{d}}}{\mathrm{argmax}} \ \mathbf{E}_{Q^{(k)}(\mathbf{M})} \left[\ln P(A^\mathcal{O},\Gamma,\mathbf{M}|\vec{\mathbf{d}})\right].
\end{cases}
\end{equation}
Next we will briefly prove that the iteration converges to the value maximizing the likelihood. 
On one hand,
\begin{align}
\ln P(A^\mathcal{O},\Gamma|\vec{\mathbf{d}}^{(k+1)})&=\ln\sum_{\mathbf{M}} P(A^\mathcal{O},\Gamma,\mathbf{M}|\vec{\mathbf{d}}^{(k+1)})\\
&\geq\sum_{\mathbf{M}} P(\mathbf{M}|A^\mathcal{O},\Gamma,\vec{\mathbf{d}}^{(k)})\ln\frac{P(A^\mathcal{O},\Gamma,\mathbf{M}|\vec{\mathbf{d}}^{(k+1)})}{P(\mathbf{M}|A^\mathcal{O},\Gamma,\vec{\mathbf{d}}^{(k)})}\\
&\geq\ln P(A^\mathcal{O},\Gamma|\vec{\mathbf{d}}^{(k)}).
\end{align}
We can see that the sequence $\{\ln P(A^\mathcal{O},\Gamma|\vec{\mathbf{d}}^{(k)})\}$  monotonously increases as $k$ grows.

On the other hand, the likelihood sequence $\{\ln P(A^\mathcal{O},\Gamma|\vec{\mathbf{d}}^{(k)})\}$ obviously has a upper bound. 
Then $\vec{\mathbf{d}}^{(k)}$ converges to the a maximum of the likelihood $P(A^\mathcal{O},\Gamma|\vec{\mathbf{d}})$~\cite{moon1996expectation}.
However, the likelihood function may have more than one local maximum values in more complex situations, while the EM algorithm is not guaranteed to converge to the global one.
To overcome the problem, we try different random initial values for the parameters repeatedly, and find the global maximum of the likelihood value when they converge~\cite{newman2016structure}.

\subsection*{Evaluation indices for reconstruction}
Accuracy, precision, recall and AUC (area under the receiver operating characteristic curve) have been widely adopted to evaluate classification methods~\cite{storey2003positive}. 
Accuracy is defined by the proportion of true results (both true positives and true negatives) among the total number of tests, i.e., accuracy = (TP+TN)/(TP+TN+FP+FN);
Precision gives the probability that a link exists in real network when reliability $Q^{\alpha}_{ij}>0.5$, i.e., precision = TP/(TP+FP);
Recall equals to the proportion of true positive rate over true positive rate and false negative rate, i.e. recall = TP/(TP+FN).
In addition, for those links whose reliability $Q^{\alpha}_{ij}=0.5$, they have half contribution to the proportion.
The area under the curve of ROC (often referred to as the AUC) quantifies the expectation that the proposed method ranks a positive one higher than a negative one.
Thus, all the tested links are ranked decreasingly according to their values of reliability, and the probability that a real link has a higher reliability than a nonexistent link is calculated. 
These four metrics are used to evaluate the proposed framework for the nine real-world data sets (Fig. S5).

\subsection*{Mean-field approximation}
Our goal in this section is to clarify the discrimination indicator introduced in Results section by mean-field approximation.
First we will calculate the entropy $\mathcal{H}(\mathbf M|A^\mathcal{O})$ determined by the ratio of the average degrees $r$, overlap of edges $v$, and percentage of partial observations $c$.
Notice that $r$ and $v$ are not available when we only have partial observations, and thus they are estimated by $\hat{r}(r,c)$ and $\hat{v}(v,c)$.
Since the average degree $\left \langle k_\alpha \right \rangle$ is a mesoscale property, we estimate $\hat{\left \langle k_\alpha \right \rangle}$ by $\mathbf{E}[\left \langle k_\alpha \right \rangle]=\sum_{\mathbf M}[Q(\mathbf{M})\cdot \left \langle k_\alpha \right \rangle{(\mathbf{M})}]$, indicating the expectation of $\left \langle k_\alpha \right \rangle$ for all potential multiplex structure.
The results shown in Fig. S6 allow us to estimate $\hat{\left \langle k_\alpha \right \rangle}$ approximately by
\begin{equation}
\begin{split}
\hat{\left \langle k_1 \right \rangle}&=\frac{\left \langle k_1 \right \rangle+\left \langle k_2 \right \rangle}{2}+[1-(1-\sqrt c)^{2/r}]\cdot\frac{\left \langle k_1 \right \rangle-\left \langle k_2 \right \rangle}{2}, \\
\hat{\left \langle k_2 \right \rangle}&=\frac{\left \langle k_1 \right \rangle+\left \langle k_2 \right \rangle}{2}-[1-(1-\sqrt c)^{2/r}]\cdot\frac{\left \langle k_1 \right \rangle-\left \langle k_2 \right \rangle}{2}. \label{meq.kc}
\end{split}
\end{equation}
Thus, we have
\begin{align}
\hat r(r,c) = \frac{\hat{\left \langle k_1 \right \rangle}}{\hat{\left \langle k_2 \right \rangle}}&=\frac{2r+(1-r)\cdot(1-\sqrt c)^{2/r}}{2-(1-r)\cdot(1-\sqrt c)^{2/r}}.
\label{eq.mrhat}
\end{align}
Noticing that $\hat r(r,0)=1$ and $\hat r(r,1)=r$, we also can approximate $\hat r$ by $\hat r\approx r^c$ for simplicity in practice. For $\hat v$, we approximate it by $\hat v\approx c\cdot v$ since $\hat v(v,0)=0$ and $\hat v(v,1)=v$.

Next, we explore the expression of entropy $\mathcal{H}$ with parameters $c$, $r$, and $v$.
Supposing that links between different nodes are independent, we obtain 
\begin{equation}
\begin{split}
\mathcal{H}(\mathbf{M}|A^\mathcal{O})&=\sum_{i=1}^N\sum_{j=1}^N\left[\mathcal{H}(M^1_{ij}|A^\mathcal{O}_{ij})+\mathcal{H}(M^2_{ij}|A^\mathcal{O}_{ij})\right].
\end{split}\label{meq.H}
\end{equation}
For the OR-aggregation mechanism, $\mathcal{H}(M^1_{ij}|A_{ij}^\mathcal{O})$ and $\mathcal{H}(M^2_{ij}|A_{ij}^\mathcal{O})$ are both equal to $0$ if $A_{ij}^\mathcal{O}=0$.
When $A_{ij}^\mathcal{O}=1$, we have
\begin{equation}
\begin{split}
\mathcal{H}(M^\alpha_{ij}|A_{ij}^\mathcal{O})&=p_\alpha\cdot\ln p_\alpha+(1-p_\alpha)\cdot\ln(1-p_\alpha),\ \alpha=1,2,
\end{split}\label{meq.H67}
\end{equation}
where 
\begin{equation}
\begin{split}
p_1=P(M^1_{ij}=1|A_{ij}^\mathcal{O}=1)=\frac{\hat v+\hat r}{1+\hat r},
\end{split}\label{meq.pa}
\end{equation}
and
\begin{equation}
\begin{split}
p_2=P(M^2_{ij}=1|A_{ij}^\mathcal{O}=1)=\frac{1+\hat v\cdot\hat r}{1+\hat r}.
\end{split}\label{meq.pb}
\end{equation}
Thus, we obtain
\begin{equation}
\begin{split}
\mathcal{H}(\mathbf M|A^\mathcal{O})=\frac{N(N-1)}{2}\cdot \left[p_1\cdot\ln p_1+(1-p_1)\cdot\ln(1-p_1)+p_2\cdot\ln p_2+(1-p_2)\cdot\ln(1-p_2)\right],
\end{split}\label{meq.H1}
\end{equation}
We empirically find that the accuracy of reconstruction is universally determined by $\mathcal{H}$ and scaling factor $\rho$, i.e., 
\begin{equation}
\begin{split}
\text{Accuracy}&\propto1-\rho\cdot \mathcal{H},
\end{split}\label{meq.acc}
\end{equation}
where
\begin{equation}
\begin{split}
\rho=\frac{1}{2\ln2\cdot N(N-1)}\cdot(1-\frac{1-v}{1+v}\cdot c^{s}).
\end{split}\label{meq.rho}
\end{equation}

Notice that
\begin{equation}
\begin{split}
P(M^2_{ij}=1|M_{ij}^1=1)=\frac{v\cdot(1+r)}{r\cdot(1+v)},
\end{split}\label{meq.p12}
\end{equation}
and
\begin{equation}
\begin{split}
P(M^1_{ij}=1|M_{ij}^2=1)=\frac{ v\cdot(1+r)}{1+v}.
\end{split}\label{meq.p21}
\end{equation}
When observing an edge in layer $\alpha$,  the probability that the edge exists in the other layer satisfies 
\begin{equation}
\begin{split}
1-P(M^\beta_{ij}=1|M_{ij}^\alpha=1)=1-\frac{\frac{ v\cdot(1+r)}{ r\cdot(1+v)}+\frac{ v\cdot(1+r)}{1+ v}/ r}{1+1/r}=\frac{1-v}{1+ v},\ \alpha,\beta\in\{ 1,2\},\ \alpha\neq\beta.
\end{split}\label{meq.pab}
\end{equation}
Thus, the term $({1-v})/({1+v})$ in  Eq. (\ref{meq.rho}) indicates that the fraction of uncertainty can be reduced by partial observations, and that $s$ describing the scale of partial observations can reduce the uncertainty of links in testing set.

\subsection*{Budget allocation}
We consider the case when we have different percentages of partial observations in each layer denoted by $c_1$ and $c_2$.
We denote $\mathcal{H}(M^1|A^\mathcal{O})$ by $\mathcal{H}_1$ for simplicity, and use $\bar c={(c_1+c_2)}/{2}$ to indicate the given budget.
Empirically, we can predict the accuracy by the function $F$ defined by 
\begin{equation}
\begin{split}
F(c_1,c_2)=  1-\frac{1-c_1}{1-c_1+(1-c_2)/ \hat r}\cdot\rho_1\cdot\mathcal{H}(M^1|A^\mathcal{O})-\frac{(1-c_2)/ \hat r}{1-c_1+(1-c_2)/ \hat r}\cdot\rho_2\cdot\mathcal{H}(M^2|A^\mathcal{O}),
\end{split}
\end{equation}
where
\begin{equation}
\begin{split}
\rho_1=\frac{1}{2\ln2\cdot N(N-1)}\cdot(1-\frac{1-v}{1+v}\cdot c_2^{s}),
\end{split}
\end{equation}
and
\begin{equation}
\begin{split}
\rho_2=\frac{1}{2\ln2\cdot N(N-1)}\cdot(1-\frac{1-v}{1+v}\cdot c_1^{s}).
\end{split}
\end{equation}
Once a certain budget ($\bar c$) is given, we regard the function $F$ as a unary function of $c_1$, i.e.,   
\begin{equation}
\begin{split}
F(c_1) = 1&-\frac{1-c_1}{1-c_1+(1-2\bar c+c_1)/ \hat r}\cdot\frac{1}{2\ln2\cdot N(N-1)}\cdot[1-\frac{1-v}{1+v}\cdot (2\bar c-c_1)^s]\cdot\mathcal{H}_1\\&-\frac{(1-2\bar c+c_1)/\hat r}{1-c_1+(1-2\bar c+c_1)/\hat r}\cdot\frac{1}{2\ln2\cdot N(N-1)}\cdot(1-\frac{1-v}{1+v}\cdot c_1^s)\cdot\mathcal{H}_2.
\end{split}
\end{equation}

We  next study the property of $F$ with $c_1$, and we will first prove that $\mathcal{H}_1\geq\mathcal{H}_2$ here.
According to the definition,
\begin{equation}
\begin{split}
\mathcal{H}_\alpha=p_\alpha\cdot\ln p_\alpha+(1-p_\alpha)\cdot\ln (1-p_\alpha),
  \ \alpha=1,2,
\end{split}\label{eq.rho2}
\end{equation}
where
\begin{equation}
\begin{split}
p_1=\frac{\hat v+\hat r}{1+\hat r},\ p_2=\frac{1+\hat v\cdot\hat r}{1+\hat r}.
\end{split}\label{eq.p1p2}
\end{equation}
Notice that the function 
\begin{equation}
\begin{split}
f(x)=x\cdot\ln x+(1-x)\cdot\ln (1-x)
\end{split}\label{eq.fx}
\end{equation}
is a monotone increasing function when $0<x\leq\frac{1}{2}$, and a monotone increasing function when $\frac{1}{2}\leq x<1$.
For $p_2$, we have
\begin{equation}
\begin{split}
p_2=\frac{1+\hat v\cdot\hat r}{1+\hat r}\geq\frac{1}{1+\hat r}\geq\frac{1}{2}.
\end{split}
\end{equation}
When $p_1\leq\frac{1}{2}$,
\begin{equation}
\begin{split}
p_2-(1-p_1)=\hat v\geq0,
\end{split}
\end{equation}
indicating $1/2\leq1-p_1\leq p_2$.
Thus, $f(1-p_1)\geq f(p_2)$, i.e., $\mathcal{H}_1\geq\mathcal{H}_2$.
When $p_1\geq\frac{1}{2}$,
\begin{equation}
\begin{split}
p_2-p_1=\frac{(1-\hat v)\cdot(1-\hat r)}{1+\hat r}\geq0,
\end{split}
\end{equation}
indicating $1/2\leq p_1\leq p_2$.
Thus, $f(p_1)\geq f(p_2)$, i.e., $\mathcal{H}_1\geq\mathcal{H}_2$.

Then, we will consider the maxima of function $F$. When $\bar c\leq1/2$ ($c_1\in[0,2\bar c]$), we have
\begin{equation}
\begin{split}
F(0) = 1-\frac{1}{2\ln2\cdot N(N-1)}\cdot\{\frac{\hat r}{\hat r+1-2\bar c}\cdot[1-\frac{1-v}{1+v}\cdot(2\bar c)^s]\cdot \mathcal{H}_1-\frac{1-2\bar c}{\hat r+1-2\bar c}\cdot\mathcal{H}_2\},
\end{split}
\end{equation}
and
\begin{equation}
\begin{split}
F(2\bar c) = 1-\frac{1}{2\ln2\cdot N(N-1)}\cdot\{\frac{\hat r(1-2\bar c)}{\hat r +1-2\hat r\bar c}\cdot\mathcal{H}_1-\frac{1}{\hat r +1-2\hat r\bar c}\cdot[1-\frac{1-v}{1+v}\cdot(2\bar c)^s]\cdot \mathcal{H}_2\}.
\end{split}
\end{equation}
Thus, we have
\begin{equation}
\begin{split}
F(0)-F(2\bar c) =\frac{ 4\hat r\bar c(\bar c-1)\cdot(\mathcal{H}_1-\mathcal{H}_2)+\frac{1-v}{1+v}\cdot(2\bar c)^s\cdot[\hat r(\hat r+1-2\bar c\hat r)\mathcal{H}_1-(r+1-2\bar c)\mathcal{H}_2]}{2\ln2\cdot N(N-1)\cdot(\hat r+1-2\bar c)\cdot(\hat r+1-2\hat r\bar c)}\leq0,
\end{split}
\end{equation}
indicating $F(0)\leq F(2\bar c)$.

When $\bar c>1/2$ ($c_1\in[2\bar c-1,1]$), we have
\begin{equation}
\begin{split}
F(2\bar c-1) =1-\frac{1}{2\ln2\cdot N(N-1)} \cdot\frac{2v}{1+v}\cdot\mathcal{H}_1,
\end{split}
\end{equation}
and
\begin{equation}
\begin{split}
F(1) =1- \frac{1}{2\ln2\cdot N(N-1)} \cdot\frac{2v}{1+v}\cdot\mathcal{H}_2,
\end{split}
\end{equation}
We have 
\begin{equation}
\begin{split}
F(2\bar c-1)-F(1) = \frac{1}{2\ln2\cdot N(N-1)} \cdot\frac{2v}{1+v}\cdot(\mathcal{H}_2-\mathcal{H}_1)\leq0,
\end{split}
\end{equation}
indicating $F(2\bar c-1)\leq F(1)$.

\subsection*{Synthetic networks generation}
To have a deep exploration of the framework proposed in this article, we generate several synthetic networks with various network characteristics for performance evaluation.
Here we mainly focus on two-layer networks with different $r$, $v$, and $\cos \left \langle \vec{d^1},\vec{d^2} \right \rangle$ as we defined in the Results section.

As shown in the Fig. 4B, we test synthetic networks ranging $0<r\leq 1$ and $0\leq v\leq1/2$. 
We  first clarify how to generate a multiple network with a given $r^*$ and $v^*$.
When given $r^*$ and $v^*$, we generate the adjacency matrix $M^1$ (the first layer in the multiplex network) by the Erdős–Rényi model, which indicates the edge $M_{ij}^1$ between any two nodes $i$ and $j$ submitted to the Bernoulli distribution
\begin{equation}
\begin{split}
P(M^1_{ij}=k)=\begin{cases} p,\text{if}\ k=1\\1-p,\ \text{if}\ k=0
\end{cases}.
\end{split} 
\end{equation}
Without loss of generality, we take $p=\frac{5}{N-1}$ such that the average degree $\left \langle k_1 \right \rangle=5$.
Then, we generate the adjacency matrix $M^2$ (the second layer in the multiplex network) by a specific way, where $M_{ij}^2$ is submitted to the distribution
\begin{equation}
\begin{split}
P(M^2_{ij}=k|M^1_{ij}=1)=\begin{cases}\frac{v^*\cdot(r^*+1)}{r^*\cdot(v^*+1)},\text{if}\ k=1\\
1-\frac{v^*\cdot(r^*+1)}{r^*\cdot(v^*+1)},\ \text{if}\ k=0
\end{cases},
\end{split} 
\end{equation}
and
\begin{equation}
\begin{split}
P(M^2_{ij}=k|M^1_{ij}=0)=\begin{cases}\frac{5}{N-6}\cdot\frac{1-v^*\cdot r^*}{r^*\cdot(v^*+1)},\text{if}\ k=1\\
1-\frac{5}{N-6}\cdot\frac{1-v^*\cdot r^*}{r^*\cdot(v^*+1)},\ \text{if}\ k=0
\end{cases}.
\end{split} 
\end{equation}
Next we  prove that the parameters $r$ and $v$ of the generated multiplex network $\mathbf{M}$ satisfy $r=r^*$ and $v=v^*$.

Obviously, the expectation for average degree $\left \langle k_1 \right \rangle$ satisfies
\begin{equation}
\begin{split}
\mathbf E(\left \langle k_1 \right \rangle)=\frac{2}{N}\cdot\frac{N\cdot(N-1)}{2}\cdot\frac{5}{N-1}=5,
\end{split} 
\end{equation}
and the expectation for average degree $\left \langle k_2 \right \rangle$ satisfies
\begin{equation}
\begin{split}
\mathbf E(\left \langle k_2 \right \rangle)&=\frac{2}{N}\cdot\frac{N\cdot(N-1)}{2}\cdot\{\frac{5}{N-1}\cdot\frac{v^*\cdot(r^*+1)}{r^*\cdot(v^*+1)}+(1-\frac{5}{N-1})\cdot\frac{5}{N-6}\cdot\frac{1-v^*\cdot r^*}{r^*\cdot(v^*+1)}\}=\frac{5}{r^*}.
\end{split} 
\end{equation}
Thus, 
\begin{equation}
\begin{split}
\mathbf E(r)=\frac{\mathbf E(\left \langle k_1 \right \rangle)}{\mathbf E(\left \langle k_2 \right \rangle)}=r^*.
\end{split} 
\end{equation}
Then, the expectation of $|E_1\cap E_2|$ satisfies
\begin{equation}
\begin{split}
\mathbf E(|E_1\cap E_2|)=\frac{5N}{2}\cdot\frac{v^*\cdot(r^*+1)}{r^*\cdot(v^*+1)},
\end{split} 
\end{equation}
and the expectation of $|E_1\cup E_2|$ satisfies
\begin{equation}
\begin{split}
\mathbf E(|E_1\cup E_2|)=\frac{5N}{2}+[\frac{N\cdot(N-1)}{2}-\frac{5N}{2}]\cdot\frac{5}{N-6}\cdot\frac{1-v^*\cdot r^*}{r^*\cdot(v^*+1)},
\end{split} 
\end{equation}
Thus, 
\begin{equation}
\begin{split}
\mathbf E(v)=\frac{\mathbf E(|E_1\cap E_2|)}{\mathbf E(|E_1\cup E_2|)}=v^*.
\end{split} 
\end{equation}

As shown in the Fig. 4C, we test a number of synthetic multiplex networks ranging $0<\cos \left \langle \vec{d^1},\vec{d^2} \right \rangle<1$.
Since $\cos \left \langle \vec{d^1},\vec{d^2} \right \rangle$ is determined by degree sequences of the two layers, we generate multiplex networks with given expectation of degree sequences.
Specifically, we first randomly generate positive vectors $\vec{d^1}$ and $\vec{d^2}$ such that the inner product ${\vec{d^1}}\cdot\vec{d^2}$ ranges from 0 to 1.
Then, we generate a number of multiplex networks with each layer being generated by the given degree sequence $\vec{d}^\alpha$.
For adjacency matrix $M^\alpha$, the element $M^\alpha_{ij}$ is submitted to a Bernoulli distribution, i.e., $P(M^\alpha_{ij}=1)=\frac{\vec{d}^\alpha(i)\cdot\vec{d}^\alpha(j)}{||\vec{d}^\alpha||_1-1}$.
This generating process can also provide multiplex networks of different $0<r_h\leq1$ as shown in the Fig. S9, A to C, since we can also generate the degree sequences with a given variance.

\subsection*{Data availability}
All data needed to evaluate the conclusions in the paper are available online as follows.
The C. elegans multiplex connectome dataset used in this study is available at \url{https://comunelab.fbk.eu/data.php}.
The London multiplex transportation network is available at \url{https://comunelab.fbk.eu/data.php}.
The temporal social interactions at the SFHH (La Société française d'Hygiène Hospitalière) conference is available at \url{www.sociopatterns.org/datasets/sfhh-conference-data-set/}.
The multiplex GPI (genetic and protein interactions) network of the Saccharomyces Pombe is available at \url{https://comunelab.fbk.eu/data.php}.
The Yeast landscape multiplex interaction networks of genes is available at \url{https://comunelab.fbk.eu/data.php}.
The multiplex air transportation network of Europe is available at \url{http://complex.unizar.es/~atnmultiplex/}.
The multiplex air transportation network of the U.S.A. is available at \url{http://stat-computing.org/dataexpo/2009/the-data.html}.
The temporal network of Wikipedia users editing each other's Talk page is available at \url{http://snap.stanford.edu/data/wiki-talk-temporal.html}.
The CollegeMsg temporal social network is available at \url{http://snap.stanford.edu/data/CollegeMsg.html}.

\subsection*{Competing Interests} The authors declare no competing financial interests.

\clearpage
\begin{figure*}
\centering
\includegraphics[width=16cm]{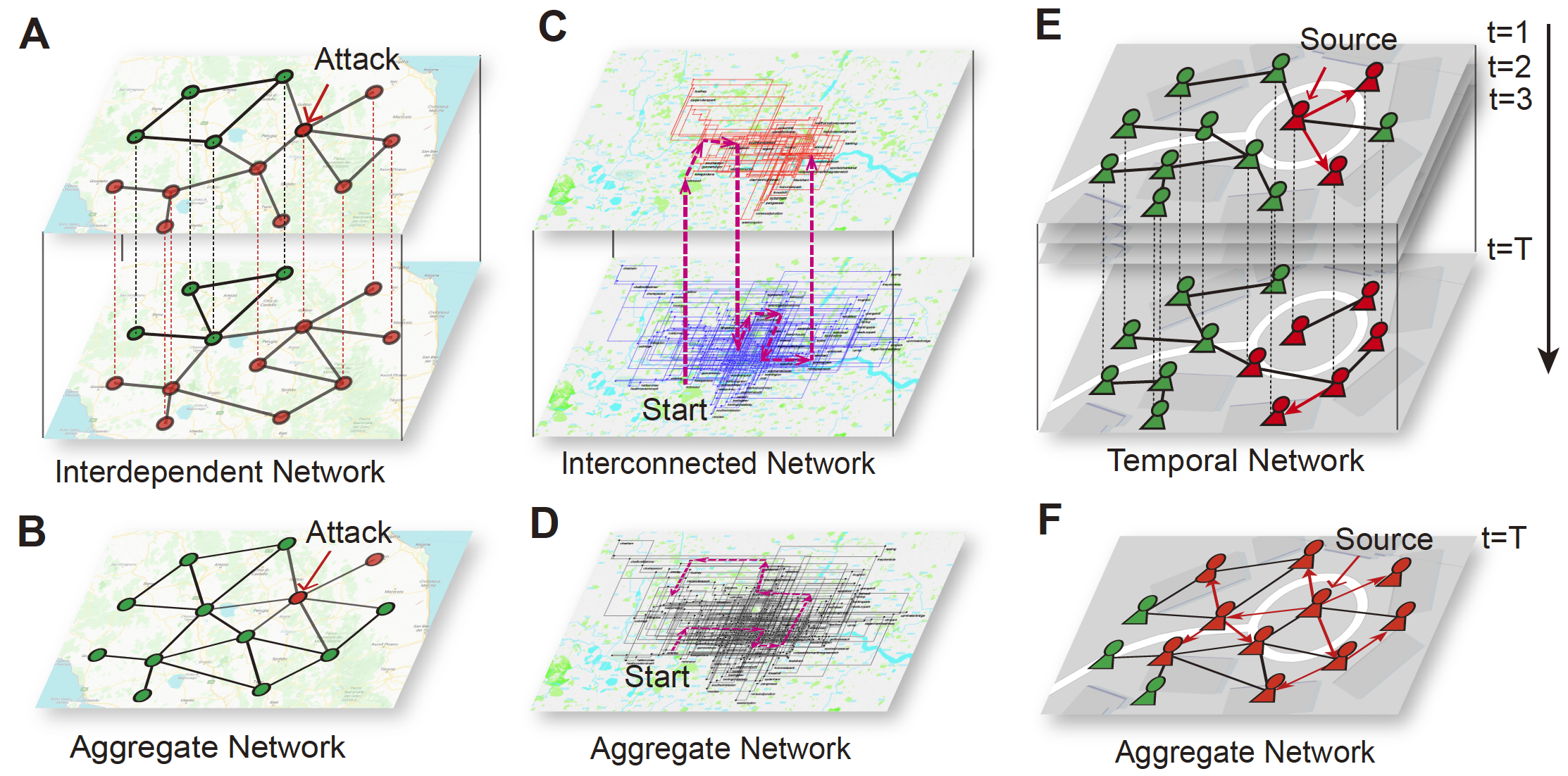}
\caption{\textbf{The different consequences for dynamics occurring on a multiplex network and the corresponding aggregate monoplex network.}
\textbf{(A)} Once a node fails due to attack, the cascade of failures will result in a catastrophic removal of nodes (red nodes in the network) in an interdependent multiplex network. 
\textbf{(B)}, Attacking the same node in the aggregate network will not trigger such a catastrophic outcome (only one node disconnects from the GCC).  
\textbf{(C)} A random walk process in a multiplex transportation network, where a walker needs to change from one layer to reach nodes in the other layer. 
\textbf{(D)} A random walk process in the corresponding aggregate network, where the walker can move along any link without any extra cost. 
\textbf{(E)} The epidemic spreads slowly, indicating less infected nodes (red nodes in the network) at $t=T$ in a temporal network since the topology changes at each time. 
\textbf{(F)} The epidemic spreads rapidly originating from the same source in the aggregate one.}
\end{figure*}

\clearpage
\begin{figure*}
\centering
\includegraphics[width=16cm]{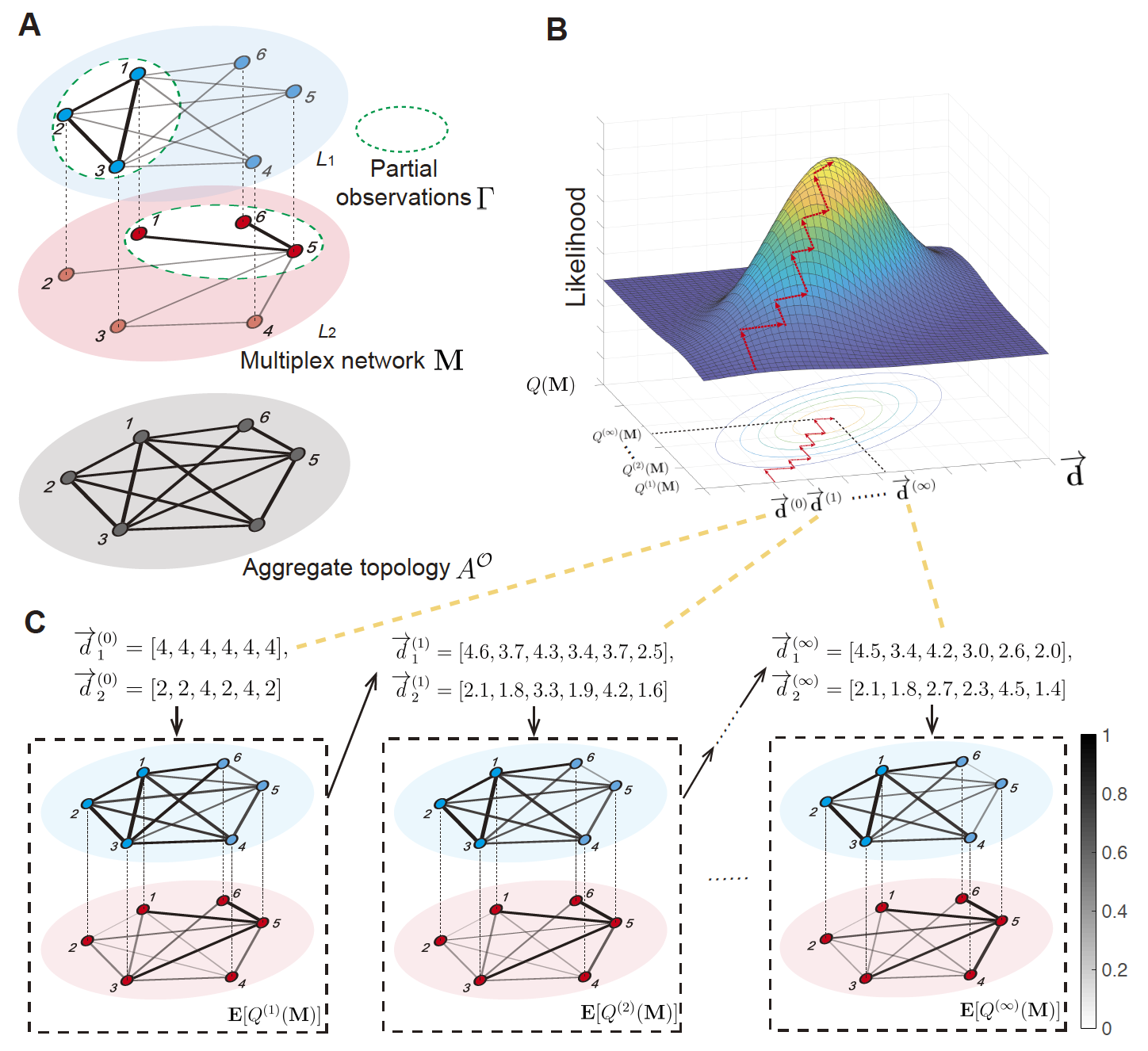}
\caption{\textbf{A schematic illustrating the reconstruction method for multiplex networks.}
\textbf{(A)} A multiplex network is aggregated to a monoplex network. 
The aggregate topology $A^\mathcal{O}$ and partial observations $\Gamma$ are leveraged to reconstruct the links in different layers that cannot be observed directly. 
\textbf{(B)} The locus of the coordinate ascent method is shown in the probability space. 
The repetitive process updating $\vec{\mathbf{d}}$ and $Q(\mathbf{M})$ maximizes the likelihood.
\textbf{(C)} A toy example is provided to demonstrate the specific steps, where the gray level of each link indicates the existent probability estimated by the proposed method.}
\end{figure*}

\clearpage
\begin{figure*}
\centering
\includegraphics[width=16cm]{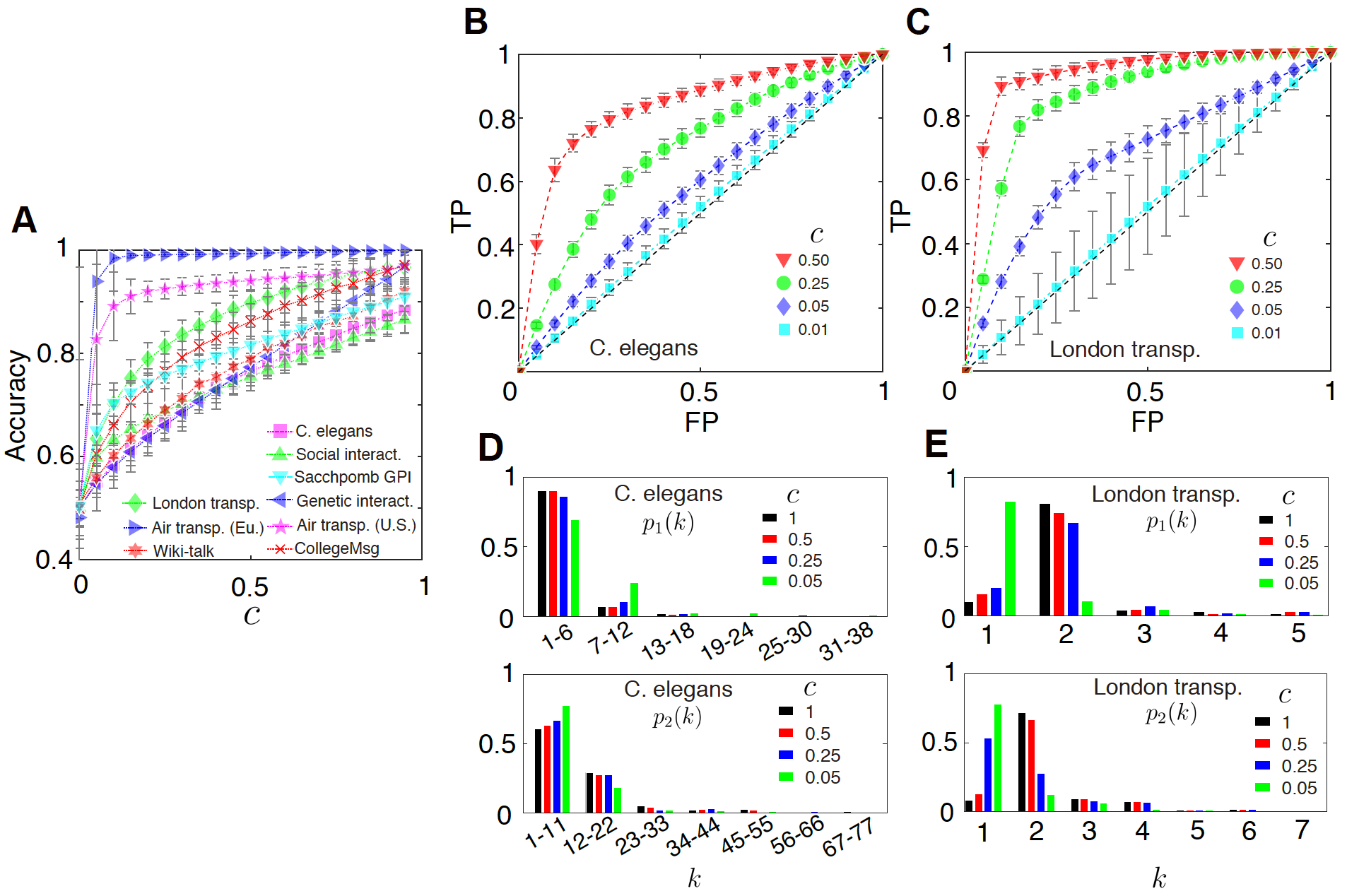}
\caption{\textbf{Performance of reconstruction in multiplex networks.}
We compare the accuracy of the reconstructed network with the proposed framework for nine real-world networks in $\textbf{(A)}$ by increasing $c$ from $0$ to $0.95$. 
The ROC space using C. elegans neural network and London transportation network is shown in $\textbf{(B)}$ and $\textbf{(C)}$, where the horizontal axis denotes the false positive rate and the vertical axis denotes the true positive rate.  
Increasing the threshold results in fewer false positives (and more false negatives), corresponding to a leftward movement on the curve from the top right corners to the left bottom along the ROC curve, where a random guess gives a point along the dashed diagonal line.
The inferred degree distributions for three values of $c$ are shown in $\textbf{(D)}$ and $\textbf{(E)}$ for the two real-world networks and compared to real ones ($c=1$). The horizontal axis indicates the degree $k$ and the vertical represents the probability that the degree of a random chosen node is equal to $k$.
}
\end{figure*}

\clearpage
\begin{figure*}
\centering
\includegraphics[width=16cm]{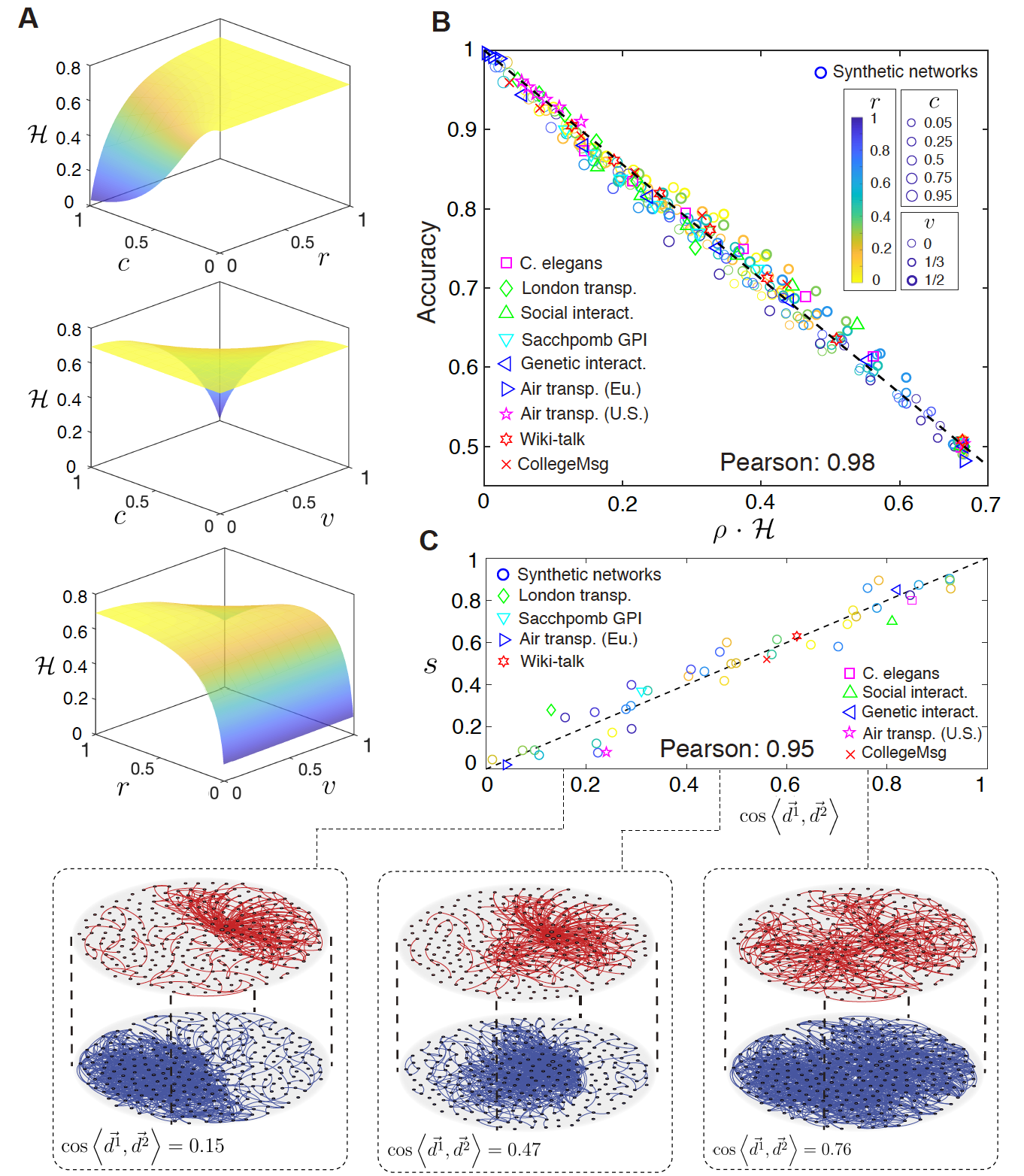}
\caption{\textbf{The impact of multiplex network characteristics on reconstruction.}
\textbf{(A)} The discrimination indicator for reconstruction is  influenced by $c$ (percentage of partial observations), $r$ (ratio of average degrees), and $v$ (overlap of edges). 
\textbf{(B)} The relationship between accuracy of reconstruction and the discrimination indicator for nine real-world networks and several synthetic networks in two dimension, showing the discrimination indicator is a good predictor (Pearson correlation is 98\%) for accuracy of reconstruction.
\textbf{(C)} The correlation between the parameter $s_{(\mathbf{M})}$ and the cosine similarity of two degree sequences $\cos \left \langle \vec{d^1},\vec{d^2} \right \rangle$ (illustrated in the lower panels) for nine real-world networks and several synthetic networks in two dimension (Pearson correlation is 95\%).
} 
\end{figure*}

\clearpage
\begin{figure*}
\centering
\includegraphics[width=16cm]{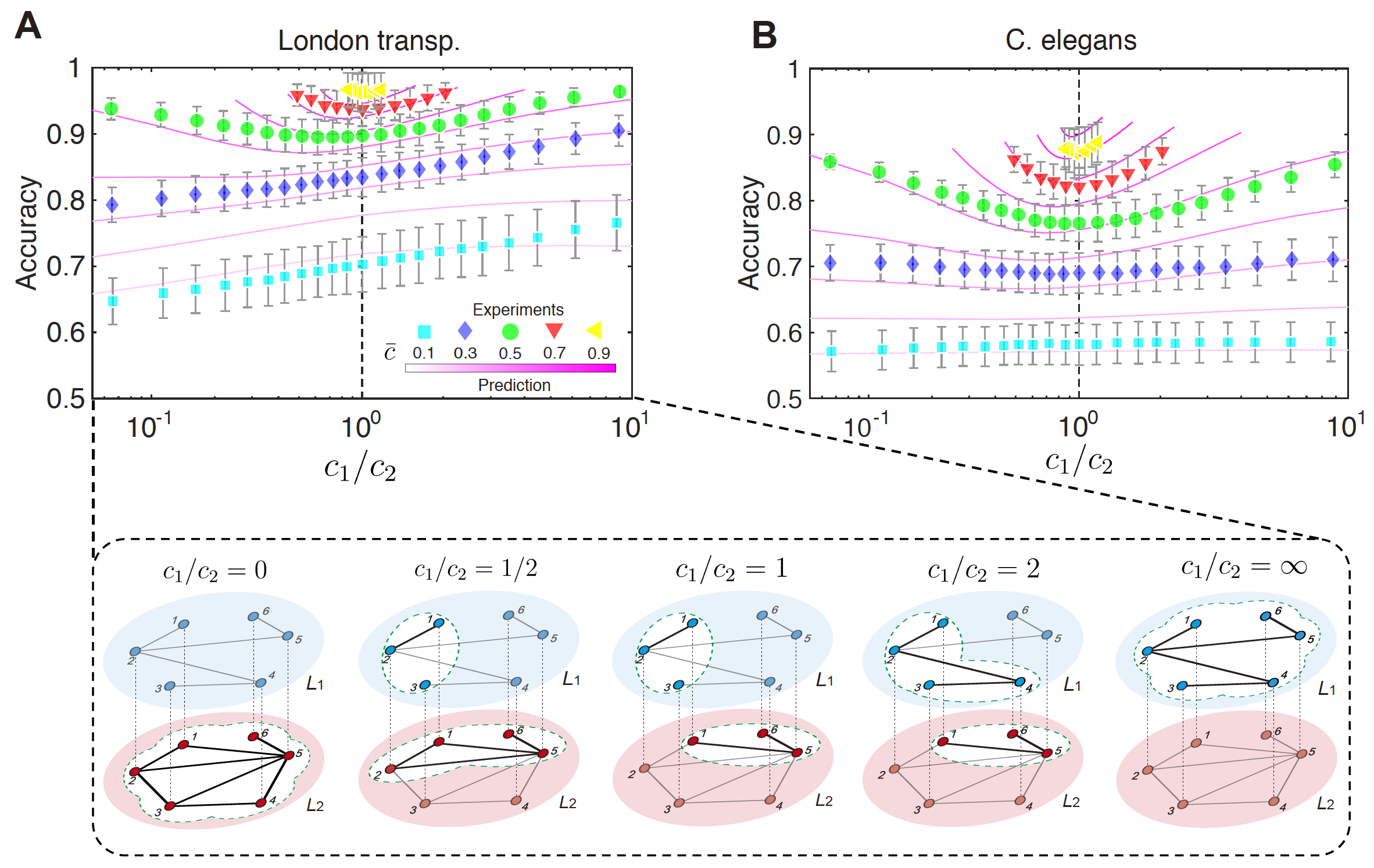}
\caption{\textbf{The allocating of budget for reconstruction.}
The accuracy of reconstruction ranging $c_1/c_2$ from 0 to $\infty$ (illustrated in the lower panels) when given total budget $\bar{c}$ for two real-world networks are shown in \textbf{(A)}, London multiplex transportation and \textbf{(B)}, C. elegans multiplex connectome. The magenta lines are the predictive results by the discrimination indicator.
}
\end{figure*}

\clearpage
\begin{figure*}
\centering
\includegraphics[width=16cm]{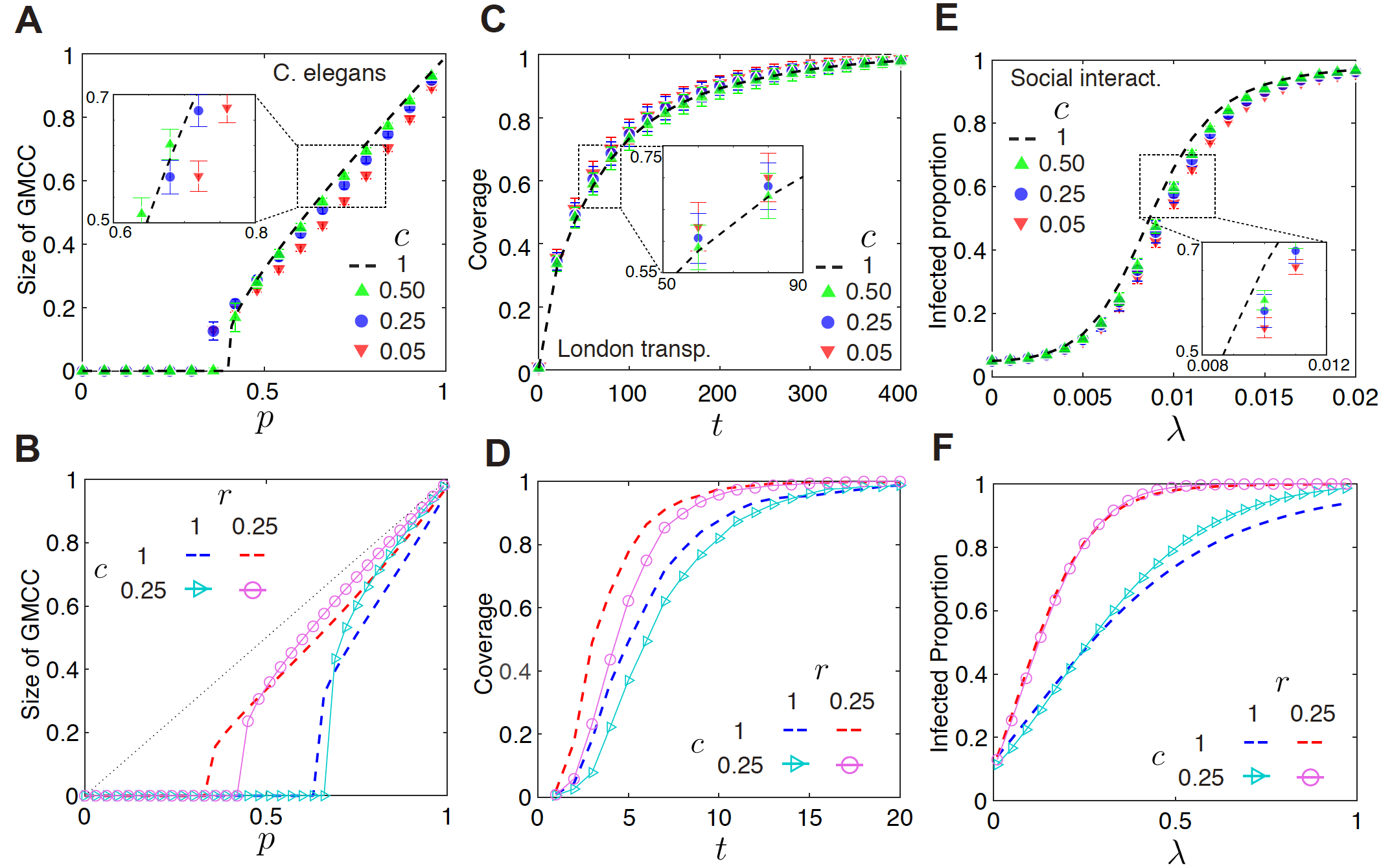}
\caption{\textbf{The performance of dynamic prediction.}
\textbf{(A)} The percolation processes of the reconstructed multiplex network $(c=0.05,0.25,0.5)$ and the real multiplex network $(c=1)$ for C. elegans multiplex connectome.
The horizontal axis denotes the occupied probability $p$ and the vertical axis denotes the size of GMCC when nodes are randomly removed with probability $1-p$ in one layer.
\textbf{(B)} The impact of $r$ on percolation process when $r=0.25$ and $r=1$.
\textbf{(C)} A random walk process taking place on the reconstructed multiplex network and real multiplex network for London transportation network.
The horizontal axis denotes  time $t$ and the vertical axis denotes coverage (the proportion of nodes that have been visited before a certain time) of $n$ walkers starting from a set of random chosen nodes.
\textbf{(D)} The impact of $r$ on random walk process when $r=0.25$ and $r=1$.
$\textbf{E}$, The spreading process on the reconstructed temporal network and real temporal network for the social interactions at the SFHH.
The horizontal axis denotes infection rate $\lambda$ and the vertical axis denotes the infected proportion.
\textbf{(F)} The impact of $r$ on spreading process when $r=0.25$ and $r=1$.}
\end{figure*}

\clearpage

\beginsupplement
\section*{Supplementary Information}\
\vspace{-40pt}
\subsection*{Supplementary Figures}\
\vspace{-10pt}
\begin{figure*}[ht]
\begin{center}
\includegraphics[width=14 cm]{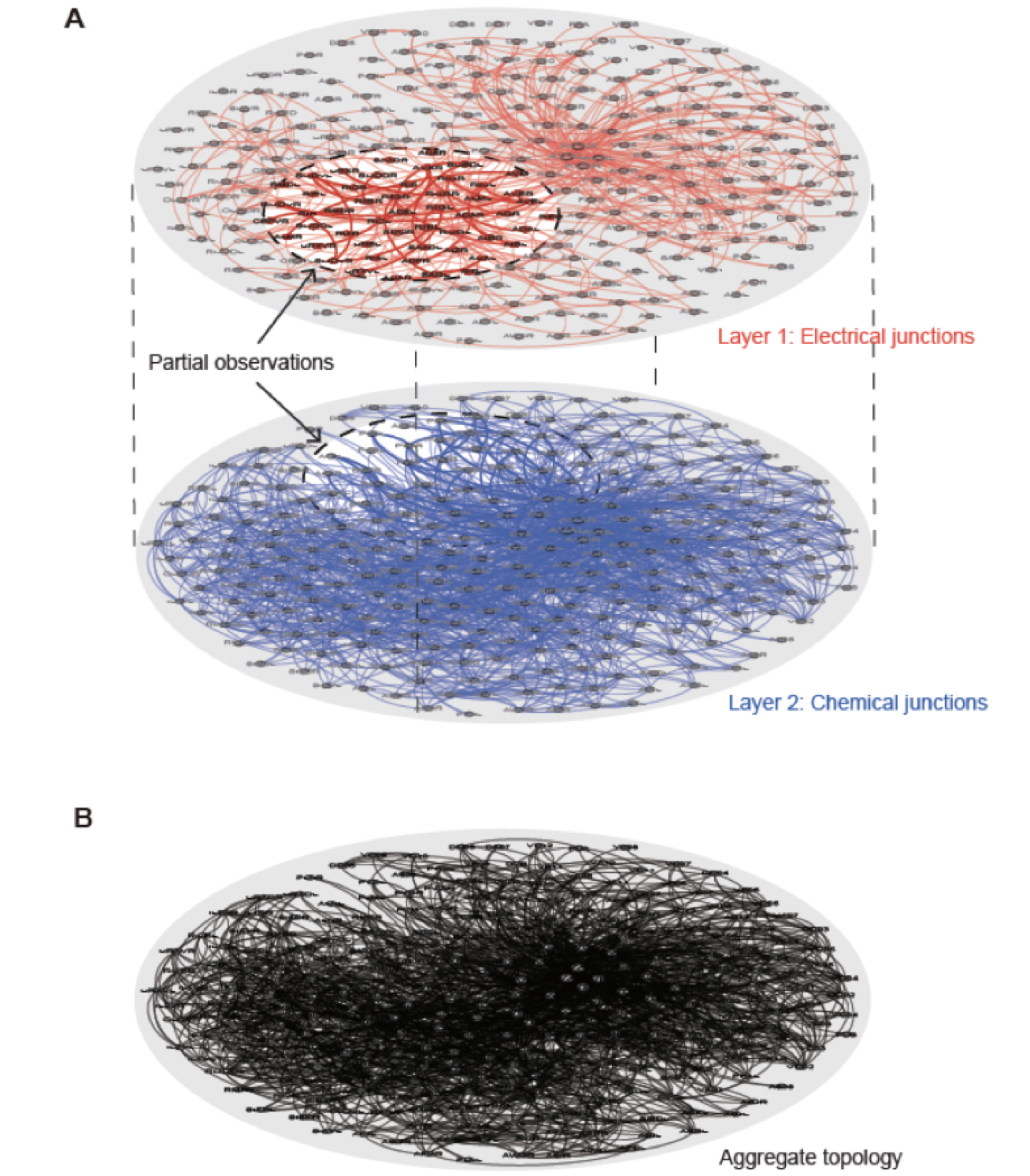}
\end{center}
\caption{\textbf{The multiplex network composed of C. elegans neuronal connectome.} 
\textbf{(A)} The multiplex network composed of two layers, indicating electrical junctions and chemical junctions, respectively. The highlighted edges in the subgraphs are the partial observations denoted by $\Gamma$. 
\textbf{(B)} The aggregate topology of the C. elegans multiplex connectome shown in \textbf{(A)}, which is a monoplex network aggregated by the OR mechanism.
}
\end{figure*}

\clearpage

\begin{figure*}[ht]
\begin{center}
\includegraphics[width=14cm]{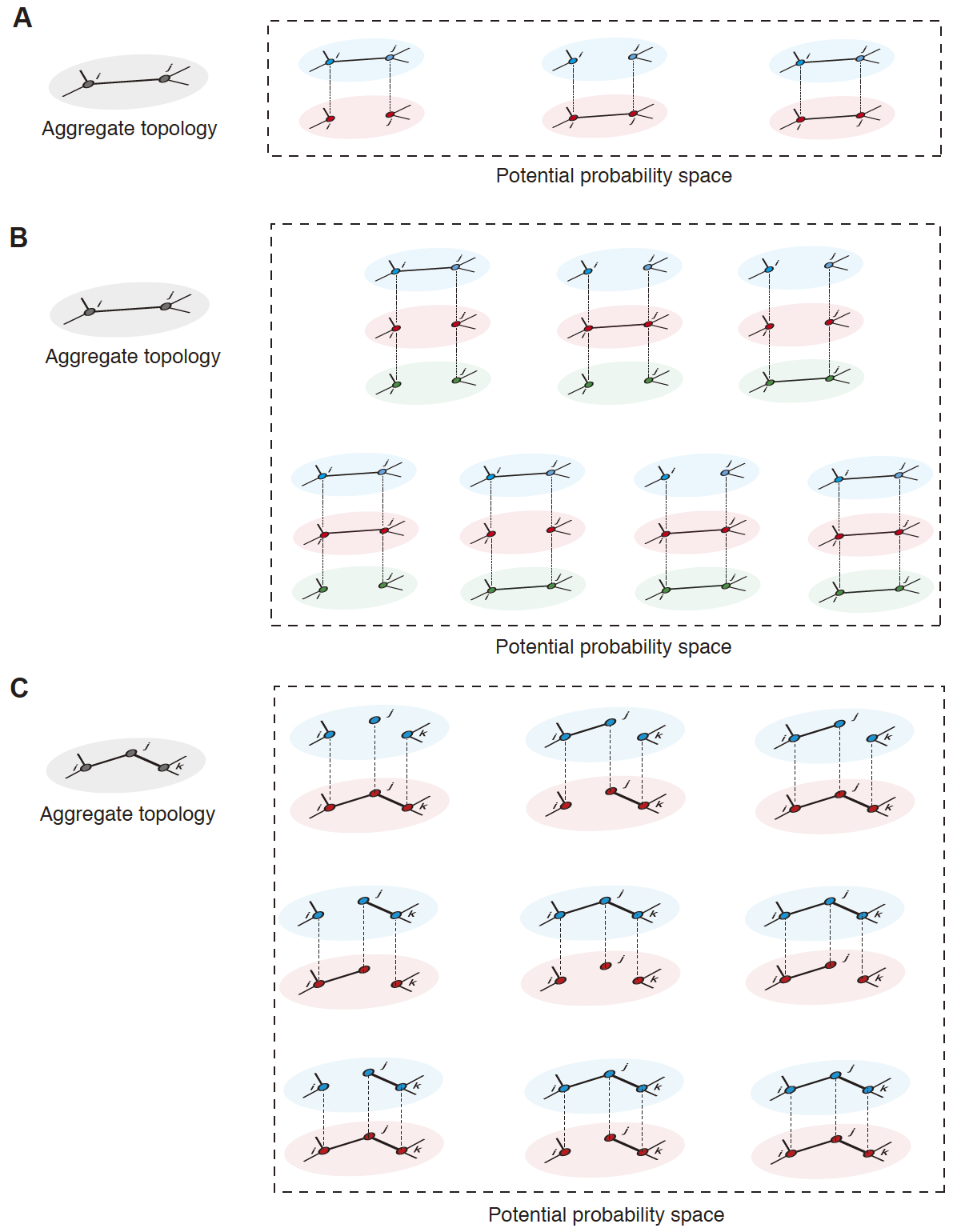}
\end{center}
\caption{\textbf{The exponential-growth probability space.} 
\textbf{(A)} When an individual link between node $i$ and node $j$ is observed in the aggregate topology, it might exist only in layer 1, only in layer 2, or in both layers, composing a three-events probability space for the multiplex structure.
\textbf{(B)} The probability space of a three-layer ($l=3$) potential multiplex structure corresponding to an individual observed link, which is composed of seven ($2^l-1$) events leading to an exponential-growth.
\textbf{(C)} Once two links between nodes $i,j$ and nodes $j,k$ are observed in the aggregate topology ($|A^\mathcal{O}|=2$), the number of potential events grows to nine ($3^{|A^\mathcal{O}|}$), resulting an exponential-growth as well. 
}
\end{figure*}

\clearpage

\begin{figure*}[ht]
\begin{center}
\includegraphics[width=8cm]{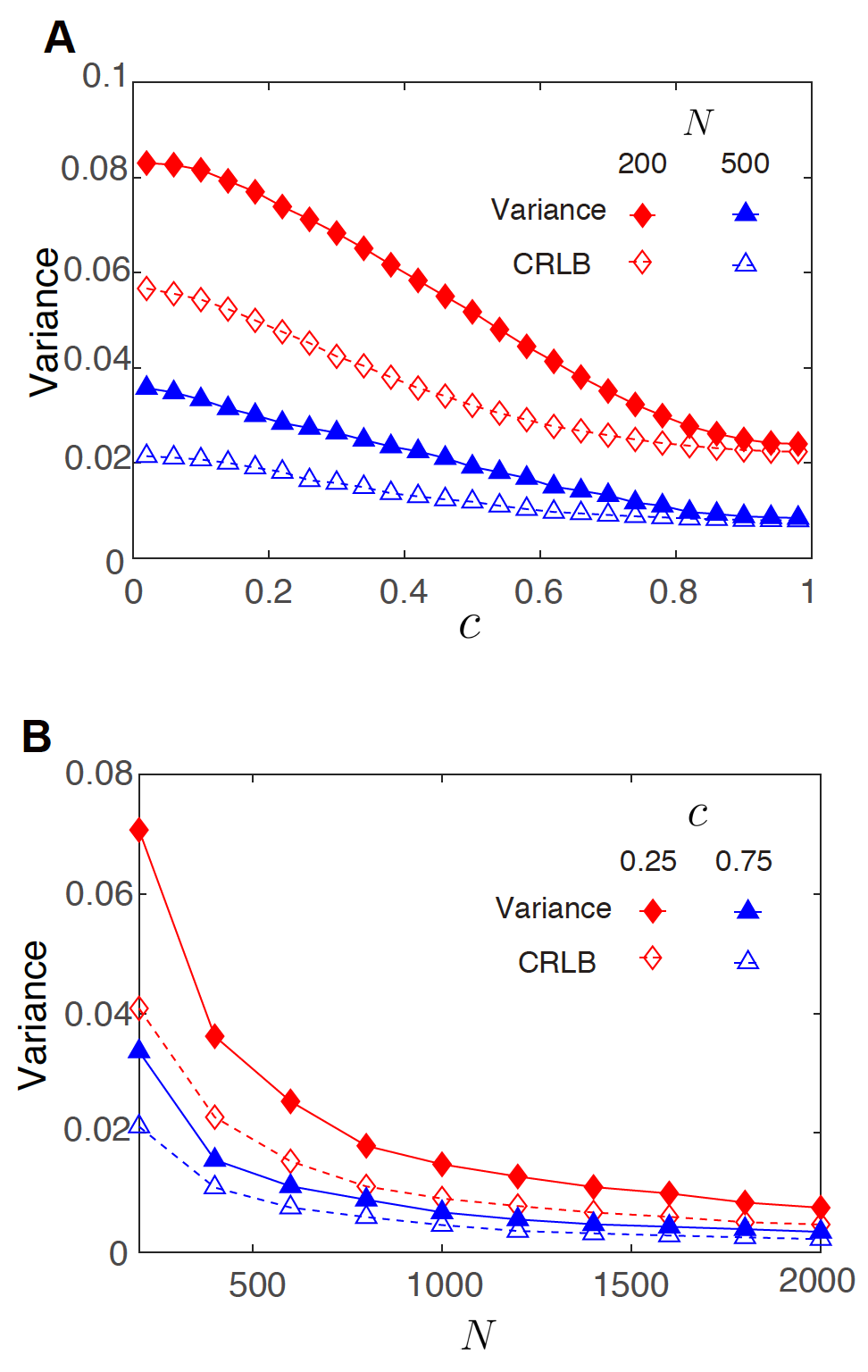}
\end{center}
\caption{\textbf{The analysis for the variance of the estimator.} The mean variances of estimated parameters and the corresponding Cramer-Rao lower bounds are shown in \textbf{(A)} and \textbf{(B)}, ranging $c$ and $N$, respectively.
These results are obtained from synthetic networks by repeating 1,000 times.}
\end{figure*}

\clearpage

\begin{figure*}[ht]
\begin{center}
\includegraphics[width=14cm]{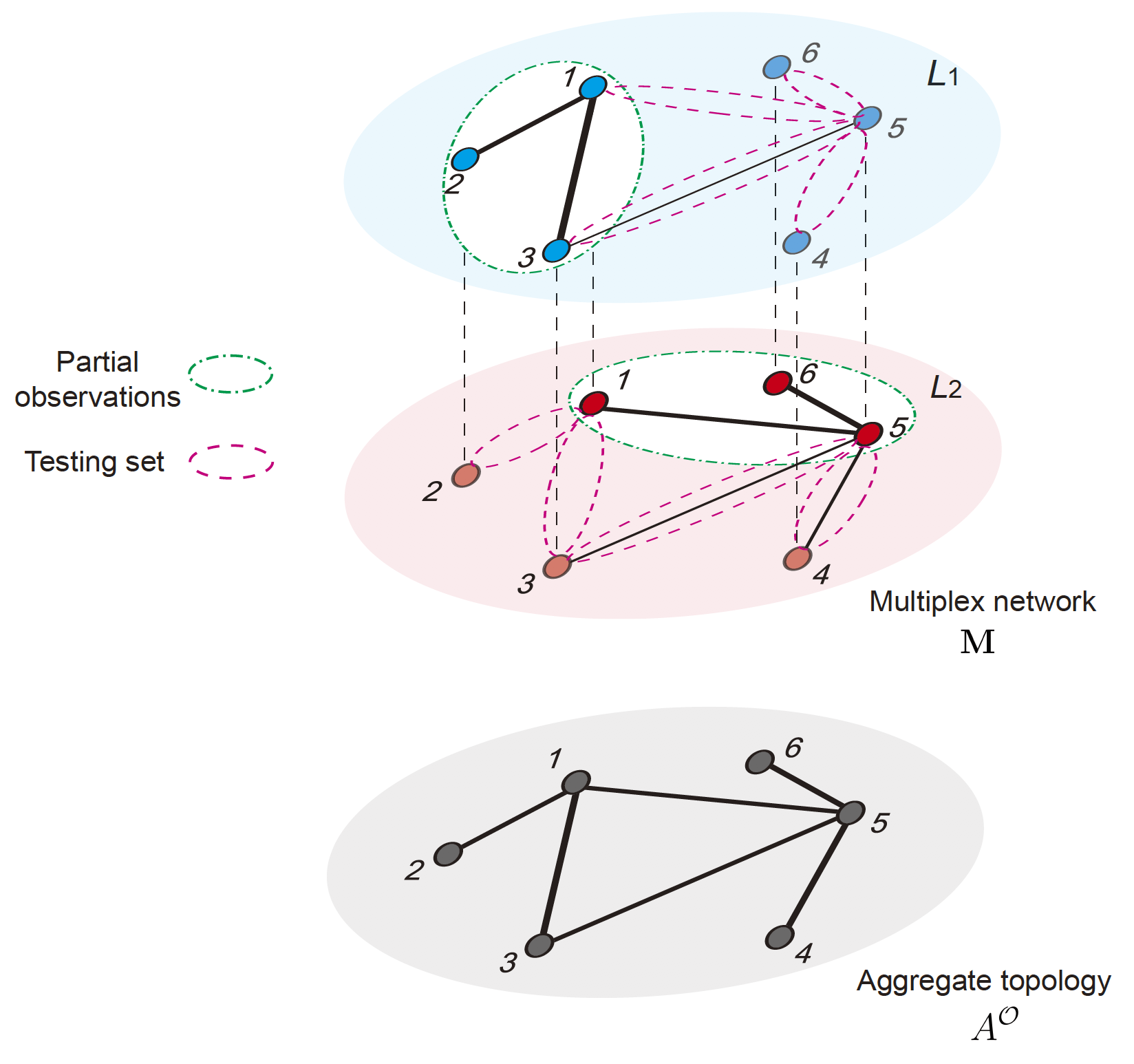}
\end{center}
\caption{\textbf{The illustration of observations and testing set.} 
The potential edges surrounded by red circles are the testing set $E^T$ consisting of potential edges except partial observations surrounded by green circles, i.e., $E^T=\{M^\alpha_{ij}\notin \Gamma|A^{\mathcal{O}}_{ij}=1\}$.
}
\end{figure*}

\clearpage

\begin{figure*}[ht]
\begin{center}
\includegraphics[width=16cm]{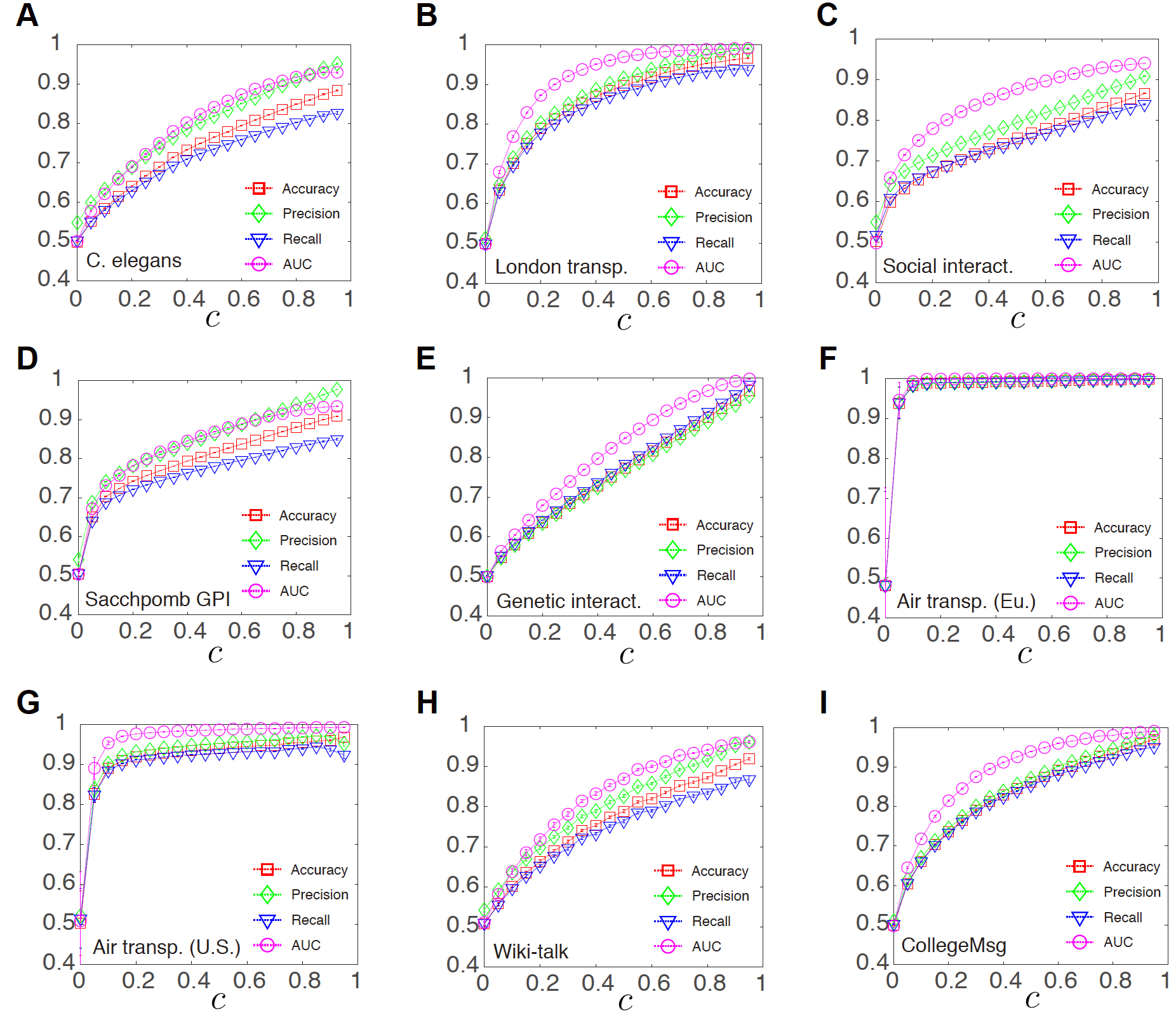}
\end{center}
\caption{\textbf{Four evaluations for reconstruction.} 
The accuracy, precision, recall and AUC are tested against $c$ (percentage of partial observations). 
These results are obtained by repeating 1,000 times from nine real-world networks: \textbf{(A)}, C. elegans connectome; \textbf{(B)}, London transportation; \textbf{(C)}, Social interaction at SFHH; \textbf{(D)}, Sacchpomb GPI; \textbf{(E)}, Genetic interaction; \textbf{(F)}, air transportation in Europe; \textbf{(G)}, air transportation in the U.S.; \textbf{(H)}, Wiki-talk network; \textbf{(I)}, CollegeMsg network.
}
\end{figure*}

\clearpage

\begin{figure*}[ht]
\begin{center}
\includegraphics[width=16cm]{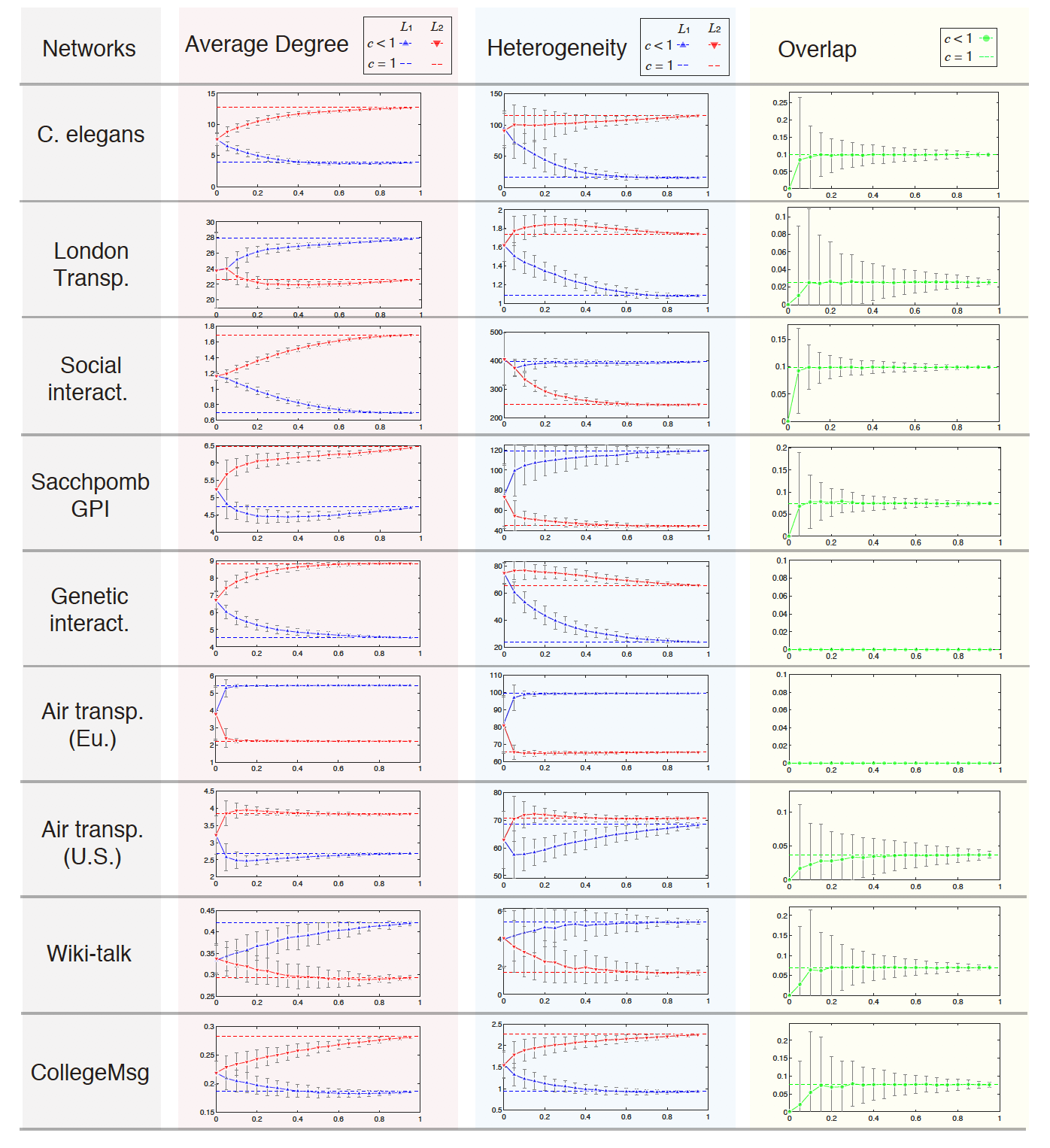}
\end{center}
\caption{\textbf{The mesoscale structure revealed in reconstructed multiplex networks.} The average degree, the heterogeneity of each layer, and the overlap of edges in the reconstructed network for nine real-world networks with $c$ increasing. }
\end{figure*}

\clearpage

\begin{figure*}[ht]
\begin{center}
\includegraphics[width=14cm]{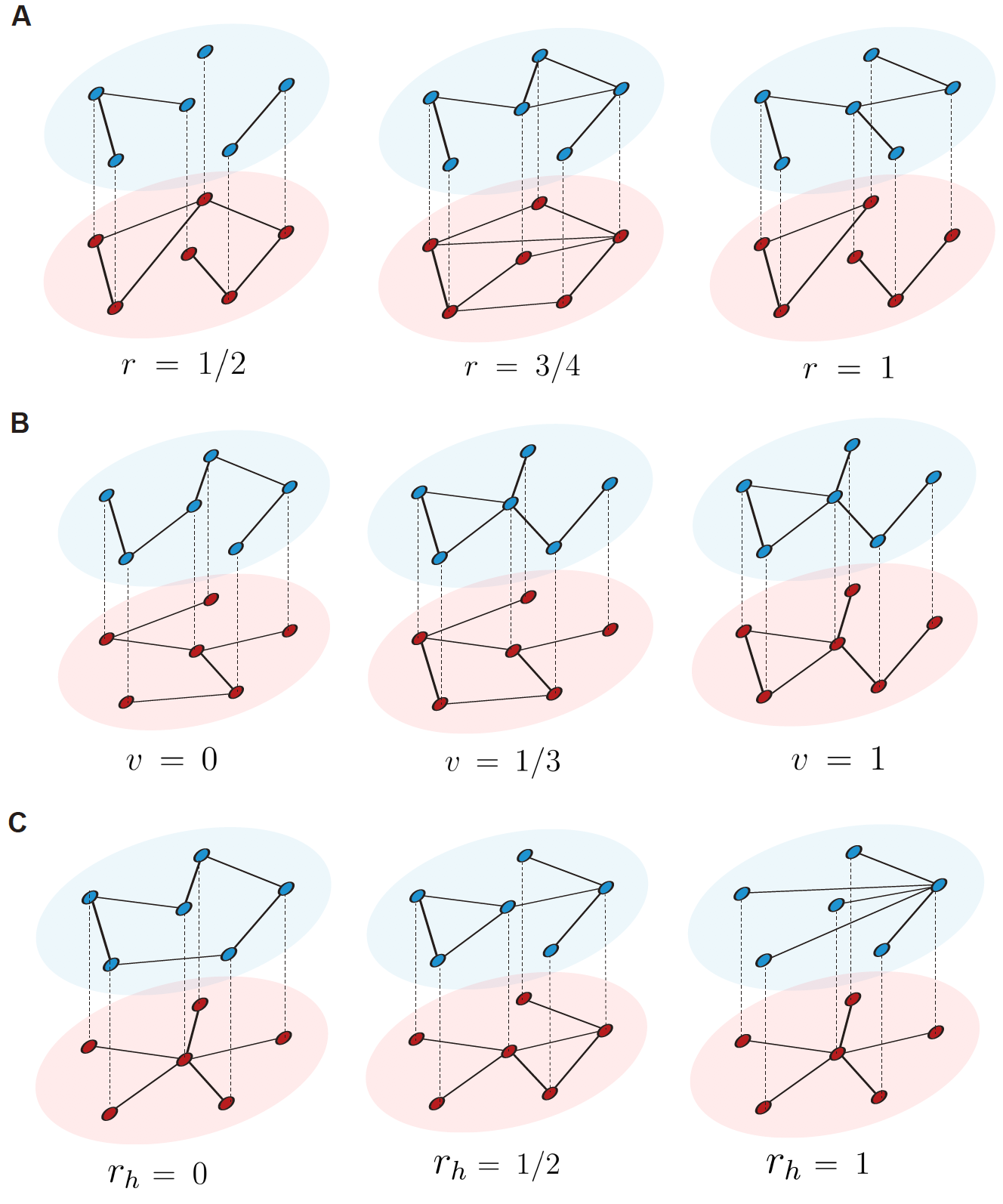}
\end{center}
\caption{\textbf{Toy examples for various multiplex network characteristics.}
We consider three main multiplex network characteristics for illustration.
Three toy examples with different characteristics are presented for \textbf{(A)}, $r={1}/{2}$,${3}/{4}$, and $1$; \textbf{(B)}, $v=0$,${1}/{3}$ and $1$; \textbf{(C)}, $r_h=0$, ${1}/{2}$, and $1$.
}
\end{figure*}

\clearpage

\begin{figure*}[ht]
\begin{center}
\includegraphics[width=14cm]{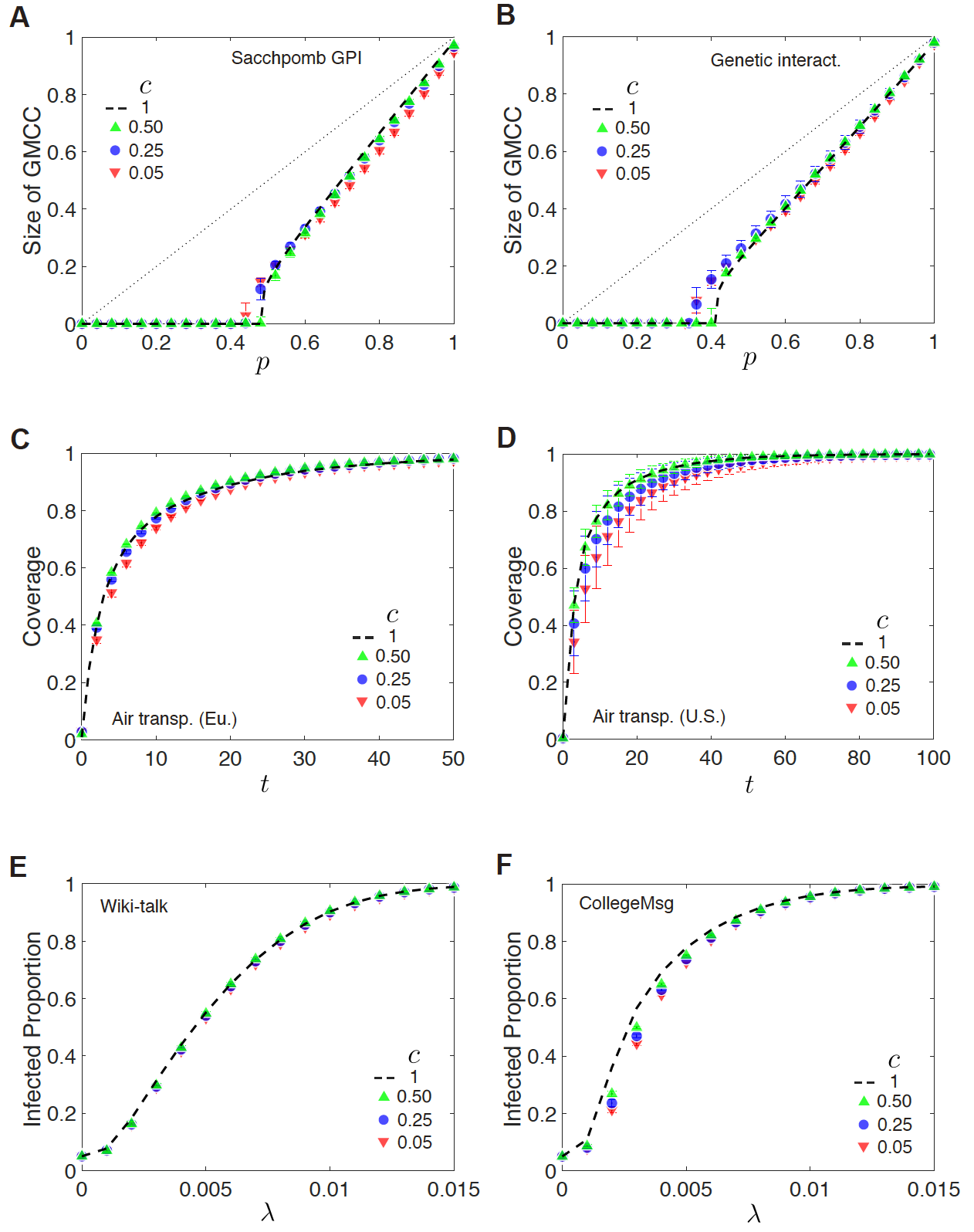}
\end{center}
\caption{\textbf{The dynamics taking place on reconstructed networks.} 
The percolation processes of the reconstructed multiplex networks (c = 0.05, 0.25, 0.5) and the real multiplex networks (c = 1) for \textbf{(A)}, Sacchpomb genetic-protein interactions and \textbf{(B)}, Yeast genetic interactions. 
The random walk process taking place on the reconstructed multiplex networks (c = 0.05, 0.25, 0.5) and real multiplex networks (c = 1) for \textbf{(C)}, air transportation networks of Europe and \textbf{(D)}, air transportation networks of the United States. 
The spreading process on the reconstructed temporal networks (c = 0.05, 0.25, 0.5) and real temporal networks (c = 1) for \textbf{(E)}, the Wiki-talk network and \textbf{(F)}, the CollegeMsg network.
}
\end{figure*}

\clearpage

\begin{figure*}[ht]
\begin{center}
\includegraphics[width=16cm]{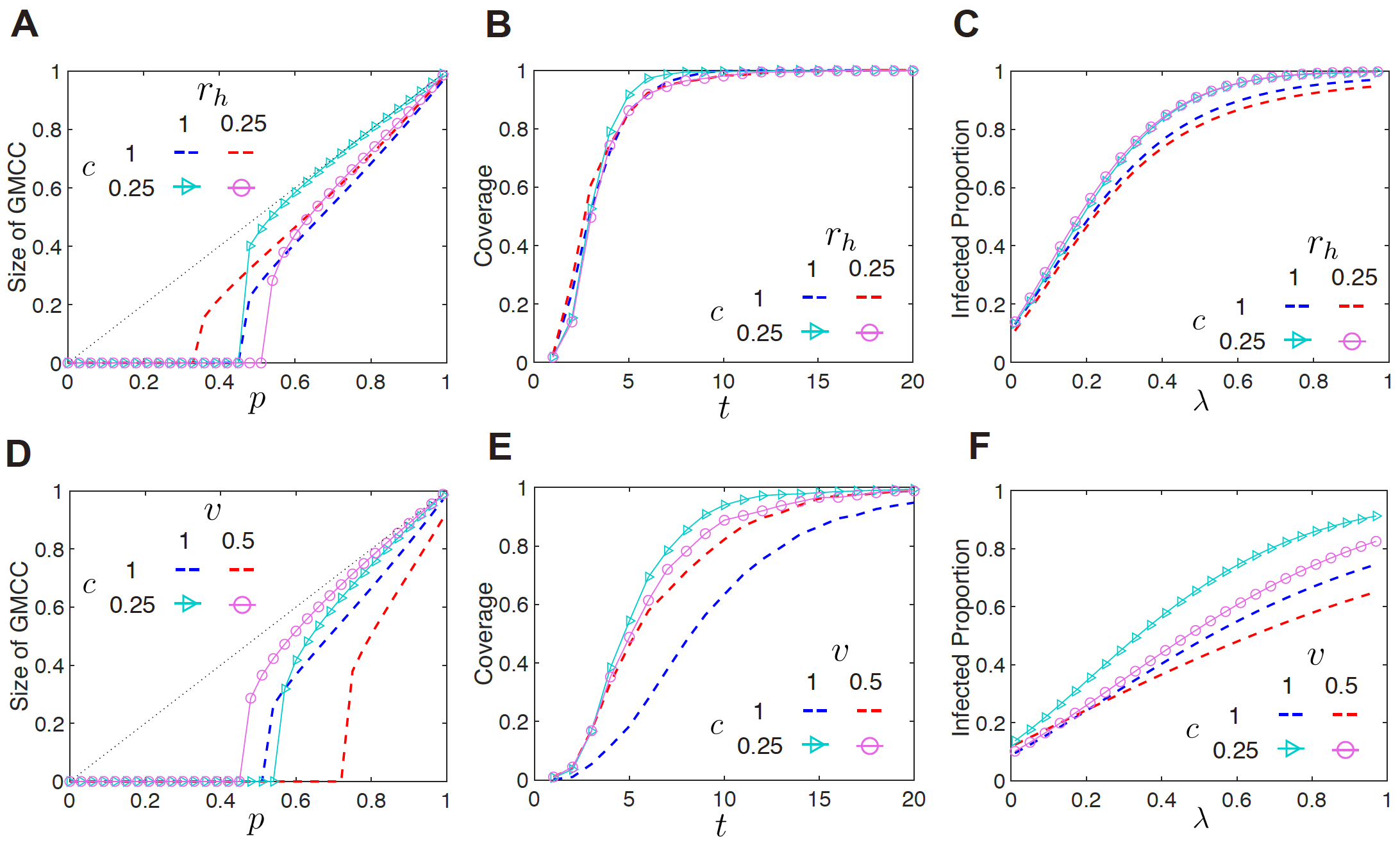}
\end{center}
\caption{\textbf{The dynamic processes influenced by various multiplex network characteristics.}
Synthetic networks shows the impact of different $r_h$ on the dynamic processes: \textbf{(A)}, percolation process; \textbf{(B)}, random walk process and \textbf{(C)}, spreading process.
Synthetic networks shows the impact of different $v$ on the dynamic processes: \textbf{(D)}, percolation process; \textbf{(E)}, random walk process and \textbf{(F)}, spreading process.
}
\end{figure*}

\clearpage

\subsection*{Supplementary Tables}\
\begin{center}
\tablefirsthead{
\hline
\multicolumn{1}{|c}{Network} &
\multicolumn{1}{c}{Layers} & 
\multicolumn{1}{c}{$N$} &
\multicolumn{1}{c}{$|E|$} &
\multicolumn{1}{c}{$\left \langle k \right \rangle$} &
\multicolumn{1}{c}{$h$} &
\multicolumn{1}{c|}{GCC} \\
\hline}
\tablehead{%
\hline
\multicolumn{7}{|l|}{\small\sl continued from previous page}\\
\hline
\multicolumn{1}{|c}{Network} &
\multicolumn{1}{c}{Layers} & 
\multicolumn{1}{c}{$N$} &
\multicolumn{1}{c}{$|E|$} &
\multicolumn{1}{c}{$\left \langle k \right \rangle$} &
\multicolumn{1}{c}{$h$} &
\multicolumn{1}{c|}{GCC} \\
\hline}
\tabletail{%
\hline
\multicolumn{7}{|r|}{\small\sl continued on next page}\\
\hline}
\tablelasttail{\hline}
\bottomcaption*{\textbf{table S1. datasets overview.} We list all multiplex networks and their properties including the number of nodes ($N$), number of edges ($|E|$), average degree ($\left \langle k \right \rangle$), variance of degree distribution ($h$) and giant connected component (GCC).}
\begin{supertabular}{|p{3.5cm}<{\centering} | p{3.5cm}<{\centering} | p{1.3cm}<{\centering} | p{1.3cm}<{\centering} | p{1.3cm}<{\centering} | p{1.3cm}<{\centering} | p{1.3cm}<{\centering} | p{1.3cm}<{\centering}}
C. elegans connectome~\cite{chen2006wiring,de2015muxviz} & Electric & 213 & 415 & 3.90 & 16.32 & 213\\
                                             & Chemical & 213 & 1353& 12.70& 115.83& 213\\
London transp.~\cite{de2014navigability} & Overground & 369 & 129 & 0.70 & 1.09 & 126\\

                                       & Underground& 369 & 312 & 1.69 & 1.074 & 271\\
Social interact. at the SFHH~\cite{Genois2018} & June 4 & 320 & 4464 & 27.90 & 396.37 & 320\\
                                 & June 5 & 320 & 3622 & 22.64 & 247.39 & 320\\
Sacchpomb GPI network~\cite{stark2006biogrid,de2015structural} & Physical association & 530 & 1254 & 4.73 & 119.01 & 530\\
         & Suppressive interaction & 530 & 1715 & 6.47 & 44.63 & 530\\
Yeast genetic interact.~\cite{costanzo2010genetic} & Positive & 506 & 1145 & 4.53 & 23.77 & 506\\
   & Negative & 506 & 2232 & 8.82 & 65.47 & 506\\
Air transp. (Eu.)~\cite{cardillo2013emergence} & Ryanair & 220 & 601 & 5.46 & 99.57 & 128\\
   & Lufthansa & 220 & 244 & 2.22 & 65.36 & 106\\
   & EasyJet & 220 & 307 & 2.79 & 46.17 & 99\\
Air transp. (U.S.) & SkyWest & 214 & 288 & 2.69 & 68.54 & 144\\
   & Southwest & 214 & 411 & 3.84 & 70.94 & 64\\
   & American Eagle & 214 & 191 & 1.79 & 40.60 & 113\\
   & American Airlines & 214 & 214 & 2.00 & 45.30 & 78\\
Social interact. (22 layers)~\cite{Genois2018} & Hour 1 & 403 & 154 & 0.76 & 6.69 & 60\\
 & Hour 2 & 403 & 144 & 0.71 & 5.54 & 37\\
 & Hour 3 & 403 & 1522 & 7.55 & 83.62 & 255\\
 & Hour 4 & 403 & 561 & 2.78 & 17.59 & 214\\
 & Hour 5 & 403 & 550 & 2.73 & 15.79 & 212\\
 & Hour 6 & 403 & 661 & 3.28 & 23.91 & 211\\
 & Hour 7 & 403 & 558 & 2.77 & 19.01 & 214\\
 & Hour 8 & 403 & 1841 & 9.14 & 78.87 & 337\\
 & Hour 9 & 403 & 861 & 4.27 & 27.69 & 276\\
 & Hour 10 & 403 & 554 & 2.75 & 10.56 & 250\\
 & Hour 11 & 403 & 136 & 0.67 & 3.06 & 66\\
 & Hour 12 & 403 & 20 & 0.10 & 0.33 & 13\\
 & Hour 13 & 403 & 31 & 0.15 & 0.55 & 24\\
 & Hour 14 & 403 & 72 & 0.36 & 1.17 & 28\\
 & Hour 15 & 403 & 193 & 0.96 & 3.81 & 98\\
 & Hour 16 & 403 & 306 & 1.52 & 7.34 & 119\\
 & Hour 17 & 403 & 3058 & 15.18 & 203.89 & 333\\
 & Hour 18 & 403 & 482 & 2.39 & 12.64 & 182\\
 & Hour 19 & 403 & 494 & 2.45 & 11.83 & 220\\
 & Hour 20 & 403 & 459 & 2.28 & 11.40 & 195\\
 & Hour 21 & 403 & 287 & 1.42 & 6.88 & 121\\
 & Hour 22 & 403 & 35 & 0.17 & 0.81 & 20\\
Wiki-talk network~\cite{paranjape2017motifs,leskovec2010governance} 
 & Week 1 & 1115 & 235 & 0.42 & 4.97 & 145\\
 & Week 2 & 1115 & 164 & 0.29 & 1.49 & 117\\
 & Week 3 & 1115 & 186 & 0.33 & 3.00 & 120\\
 & Week 4 & 1115 & 161 & 0.29 & 1.71 & 102\\
 & Week 5 & 1115 & 249 & 0.45 & 3.20 & 138\\
 & Week 6 & 1115 & 190 & 0.34 & 1.93 & 119\\
 & Week 7 & 1115 & 301 & 0.54 & 6.96 & 167\\
 & Week 8 & 1115 & 258 & 0.46 & 4.85 & 168\\
 & Week 9 & 1115 & 335 & 0.60 & 7.35 & 194\\
 & Week 10 & 1115 & 362 & 0.65 & 6.26 & 213\\
 & Week 11 & 1115 & 334 & 0.60 & 5.60 & 187\\
 & Week 12 & 1115 & 414 & 0.74 & 7.82 & 235\\
 & Week 13 & 1115 & 428 & 0.77 & 8.16 & 223\\
 & Week 14 & 1115 & 439 & 0.79 & 10.55 & 274\\
CollegeMsg network~\cite{panzarasa2009patterns}
 & Day 1 & 1209 & 113 & 0.19 & 0.61 & 78\\
 & Day 2 & 1209 & 171 & 0.28 & 1.46 & 115\\
 & Day 3 & 1209 & 164 & 0.27 & 1.06 & 107\\
 & Day 4 & 1209 & 145 & 0.24 & 0.66 & 101\\
 & Day 5 & 1209 & 268 & 0.44 & 1.80 & 149\\
 & Day 6 & 1209 & 250 & 0.41 & 1.32 & 156\\
 & Day 7 & 1209 & 327 & 0.54 & 1.99 & 182\\
 & Day 8 & 1209 & 426 & 0.70 & 3.97 & 222\\
 & Day 9 & 1209 & 333 & 0.55 & 2.00 & 201\\
 & Day 10 & 1209 & 275 & 0.45 & 1.64 & 183\\
 & Day 11 & 1209 & 406 & 0.67 & 2.69 & 231\\
 & Day 12 & 1209 & 630 & 1.04 & 5.94 & 325\\
 & Day 13 & 1209 & 583 & 0.96 & 4.01 & 310\\
 & Day 14 & 1209 & 622 & 1.03 & 4.74 & 330\\
 & Day 15 & 1209 & 700 & 1.16 & 5.41 & 356\\
 & Day 16 & 1209 & 676 & 1.12 & 17.15 & 436\\
 & Day 17 & 1209 & 399 & 0.66 & 2.46 & 263\\
 & Day 18 & 1209 & 576 & 0.95 & 4.31 & 333\\
 & Day 19 & 1209 & 233 & 0.39 & 1.14 & 146\\
 & Day 20 & 1209 & 566 & 0.94 & 3.18 & 359\\
 & Day 21 & 1209 & 508 & 0.84 & 2.73 & 341\\
 & Day 22 & 1209 & 117 & 0.19 & 0.29 & 77\\
 & Day 23 & 1209 & 439 & 0.73 & 3.87 & 307\\
 & Day 24 & 1209 & 326 & 0.54 & 1.65 & 243\\
 & Day 25 & 1209 & 468 & 0.77 & 2.82 & 305\\
 & Day 26 & 1209 & 613 & 1.01 & 7.63 & 392\\
 & Day 27 & 1209 & 530 & 0.88 & 2.64 & 342\\
 & Day 28 & 1209 & 579 & 0.96 & 3.50 & 351\\
\end{supertabular}
\end{center}
\clearpage

\begin{center}
\tablefirsthead{%
\hline
\multicolumn{1}{|c}{Network} & 
\multicolumn{1}{c}{Layers} &
\multicolumn{1}{c}{$\cos \left \langle \vec{d^1},\vec{d^2} \right \rangle$} &
\multicolumn{1}{c|}{$s_{(\mathbf M)}$} \\
\hline}
\tablehead{%
\hline
\multicolumn{4}{|l|}{\small\sl continued from previous page}\\
\hline
\multicolumn{1}{|c}{Network} & 
\multicolumn{1}{c}{Layers} &
\multicolumn{1}{c}{$\cos \left \langle \vec{d^1},\vec{d^2} \right \rangle$} &
\multicolumn{1}{c|}{$s_{(\mathbf M)}$} \\
\hline}
\tabletail{%
\hline
\multicolumn{4}{|r|}{\small\sl continued on next page}\\
\hline}
\tablelasttail{\hline}
\bottomcaption*{\textbf{table S2. cosine similarity and $s$ of each multiplex network.} We list all datasets tested in the analysis of entropy, and their properties including cosine similarity $\cos \left \langle \vec{d^1},\vec{d^2} \right \rangle$ and $s_{(\mathbf M)}$.}
\begin{supertabular}{|p{5cm}<{\centering} | p{5cm}<{\centering} | p{2cm}<{\centering} | p{2cm}<{\centering} |} 
C. elegans  & Electric  & 0.85 & 0.80 \\
 & Chemical & & \\
London transp. & Overground & 0.13 & 0.28 \\
 & Underground & & \\
Social interact. & June 4 & 0.81 & 0.70 \\
 & June 5 & & \\
Sacchpomb GPI network & Physical association & 0.31 & 0.37 \\
 & Suppressive interaction & & \\
Yeast genetic interact. & Positive & 0.82 & 0.85 \\
 & Negative & & \\
Air transp. (Eu.) & Ryanair & 0.04 & 0.02 \\
 & Lufthansa & & \\
Air transp. (U.S.) & SkyWest & 0.24 & 0.08 \\
 & Southwest & & \\
Wiki-talk network  & Week 1 & 0.62 & 0.63 \\
 & Week 2 & & \\
CollegeMsg network & Day 1 & 0.56 & 0.52 \\
 & Day 2 & & \\
\end{supertabular}
\end{center}

\clearpage
\subsection*{Supplementary Text}
\subsection*{A$\ \ \ $ Aggregate mechanisms}
There are many aggregate mechanisms for mapping a multiplex network $\mathbf{M}$ to a monoplex network $A^\mathcal{O}$. 
Here we list three common cases. 
Observing the aggregation with logical relationship ``OR'' is the most common mechanism in real life. 
We adopt this aggregate mechanism in this article for illustration, and denote by $\varphi_{\text{OR}}$  the mapping with relationship ``OR''.  
Then we have
\begin{equation*}
\begin{split}
A^\mathcal{O}=\varphi_{\text{OR}}(\mathbf{M})=\mathbf{1}_{N\times N}-\prod^M_{\alpha=1} (\mathbf{1}_{N\times N}-M^{\alpha})
\end{split},
\end{equation*}
where $\mathbf{1}_{N\times N}$ is the matrix with all elements equaling to one, and
\begin{equation*}
\begin{split}
A_{ij}^\mathcal{O}=\varphi_{\text{OR}}(\mathbf{M}_{ij})=1-\prod^M_{\alpha=1} (1-M_{ij}^{\alpha})
\end{split}.
\end{equation*}
The ``OR'' aggregate mechanism maps an unweighted multiplex network to an unweighted monoplex network. 
For example, assuming that the $\alpha$-th layer $M^{\alpha}$ is an undirected, unweighted network generated by the ER network model with parameter $\theta^{\alpha}$, the distribution of a link $A_{ij}^{\mathcal{O}}$ in the aggregated network $A^{\mathcal{O}}$ with ``OR'' is submitted to a Bernoulli distribution
\begin{equation*}
P(A^{\mathcal{O}}_{ij}=k)=
\begin{cases}
1-\prod_\alpha(1-\theta^{\alpha}),& \text {if}\ k=1\\
\prod_\alpha(1-\theta^{\alpha}),& \text {if}\ k=0
\end{cases}.
\end{equation*}
Then, the joint distribution of observing the whole aggregated network is 
\begin{equation*}
P(A^{\mathcal{O}}|\Theta)=\prod_{i<j}\left[1-\prod_\alpha(1-\theta^{\alpha})\right]^{A^{\mathcal{O}}_{ij}}\cdot\left[\prod_\alpha(1-\theta^{\alpha})\right]^{1-A^{\mathcal{O}}_{ij}}.
\end{equation*}

The second mechanism obtains the aggregated network by simple aggregation with summation, i.e., 
\begin{equation*}
\begin{split}
A^\mathcal{O}=\varphi_{\text{SUM}}(\mathbf{M})=\sum^L_{\alpha=1}M^{\alpha}
\end{split},
\end{equation*}
and
\begin{equation*}
\begin{split}
A_{ij}^\mathcal{O}=\varphi_{\text{SUM}}(\mathbf{M}_{ij})=\sum^L_{\alpha=1}M_{ij}^{\alpha}
\end{split}.
\end{equation*}

This aggregation  can map an unweighted multiplex network to a weighted monoplex network. 
Specifically, if $M^\alpha$ is an unweighted multiplex network, i.e., $M_{ij}^{\alpha}\in\{0,1\}$,  the aggregate network may not be unweighted network any longer, because $A_{ij}^\mathcal{O}\in\{0,1,2,\cdots,L\}$.  
For example, suppose that the $\alpha$-th layer $M^{\alpha}$ is an undirected, unweighted network generated by the ER network model with parameter $\theta$, i.e., 
\begin{equation*}
P(M^{\alpha}_{ij}=k)=
\begin{cases}
\ \ \ \theta\ \ ,& \text{if}\ k=1\\
1-\theta,& \text{if}\ k=0
\end{cases}.
\end{equation*}
The distribution of an individual link $A_{ij}^{\mathcal{O}}$ in the aggregate network $A^{\mathcal{O}}$ with the ``SUM'' mechanism is submitted to a multinomial distribution
\begin{equation*}
P(A^{\mathcal{O}}_{ij}=k)=\binom{L}{k}\cdot\theta^k(1-\theta)^{L-k},\ k=0,1,2,\cdots,L.
\end{equation*}
Then, we can obtain the joint distribution of the whole aggregate network with all links  
\begin{equation*}
P(A^{\mathcal{O}}|\theta)=\prod_{i<j}\left[\binom{L}{A^{\mathcal{O}}_{ij}}\cdot\theta^{A^{\mathcal{O}}_{ij}}(1-\theta)^{L-A^{\mathcal{O}}_{ij}}\right].
\end{equation*}

The logical aggregate mechanism ``AND'' is also common in real life, and we denote the mapping with logical relationship ``AND'' by $\varphi_{\text{AND}}$. 
Thus, we have
\begin{equation*}
\begin{split}
A^\mathcal{O}=\varphi_{\text{AND}}(\mathbf{M})=\prod^L_{\alpha=1} M^{\alpha}
\end{split},
\end{equation*}
and its elements can be specified by
\begin{equation*}
\begin{split}
A_{ij}^\mathcal{O}=\varphi_{\text{AND}}(\mathbf{M}_{ij})=\prod^L_{\alpha=1} M_{ij}^{\alpha}
\end{split}.
\end{equation*}
The ``AND'' aggregation will also map an unweighted multiplex network to an unweighted monoplex network. 
For example, assuming that the $\alpha$-th layer $M^{\alpha}$ is an undirected, unweighted network generated by the ER network model with parameter $\theta^{\alpha}$, the distribution of an individual link $A_{ij}^{\mathcal{O}}$ in the aggregate network $A^{\mathcal{O}}$ with ``AND'' is submitted to a Bernoulli distribution
\begin{equation*}
P(A^{\mathcal{O}}_{ij}=k)=
\begin{cases}
\prod_\alpha\theta^{\alpha},& \text{if}\ k=1\\
1-\prod_\alpha\theta^{\alpha},& \text{if}\ k=0
\end{cases}.
\end{equation*}
Then, we have the joint distribution of the aggregate network 
\begin{equation*}
P(A^{\mathcal{O}}|\Theta)=\prod_{i<j}\left[\prod_\alpha\theta^{\alpha}\right]^{A^{\mathcal{O}}_{ij}}\cdot\left[1-\prod_\alpha\theta^{\alpha}\right]^{1-A^{\mathcal{O}}_{ij}}.
\end{equation*}

\subsection*{B$\ \ \ $ Complete algorithms}
In this section we will present complete algorithms for the specific cases in simulations, and we suppose $L=2$ for simplicity.
Notice that in the proposed method, the likelihood
\begin{equation}
\begin{split}
P(A^\mathcal{O},\Gamma|\vec{d}^{1},\vec{d}^{2})
&=\sum_{M^{1},M^{2}}P(A^\mathcal{O},\Gamma,M^{1},M^{2}|\vec{d}^{1},\vec{d}^{2}).
\end{split}
\end{equation}
We thus have
\begin{equation}
\begin{split}
\ln{P(A^\mathcal{O},\Gamma|\vec{d}^{1},\vec{d}^{2})}\geq\sum_{M^{1},M^{2}}Q(M^{1},M^{2})\ln\frac{P(A^\mathcal{O},\Gamma,M^{1},M^{2}|\vec{d}^{1},\vec{d}^{2})}{Q(M^{1},M^{2})},
\end{split}\label{si.ji}
\end{equation}
where we employing the Jensen's inequality in the above.
Notice that in the Jensen's inequality Eq. (\ref{si.ji}), the equality holds if and only if
\begin{equation}
\begin{split}
Q(M^{1},M^{2})&=\frac{P(A^\mathcal{O},\Gamma,M^{1},M^{2}|\vec{d}^{1},\vec{d}^{2})}{\sum\limits_{M^{1},M^{2}}P(A^\mathcal{O},\Gamma,M^{1},M^{2}|\vec{d}^{1},\vec{d}^{2})}\\
&=\frac{P(A^\mathcal{O},\Gamma,M^{1},M^{2}|\vec{d}^{1},\vec{d}^{2})}{P(A^\mathcal{O},\Gamma|\vec{d}^{1},\vec{d}^{2})}\\
&=P(M^{1},M^{2}|A^\mathcal{O},\Gamma,\vec{d}^{1},\vec{d}^{2}).
\end{split}
\end{equation}
However, the parameters $\vec{d}^{1},\vec{d}^{2}$ and the probability distribution $Q(M^{1},M^{2})$ are interdependent. 

We denote 
\begin{equation}
\begin{split}
&J(Q,\vec{\mathbf d})=\sum_{M^{1},M^{2}}Q(M^{1},M^{2})\ln\frac{P(A^\mathcal{O},\Gamma,M^{1},M^{2}|\vec{d}^{1},\vec{d}^{2})}{Q(M^{1},M^{2})},
\end{split}\label{eq.lb}
\end{equation}
indicating $J$ is a function of distribution $Q$ and parameters $\vec{\mathbf d}$ (i.e., $\vec{d}^{1}$ and $\vec{d}^{2}$).
Thus, in the E-step, we maximize the function $J(Q, \vec{\mathbf d})$ with respect to the distribution $Q$ while keeping $\vec{d}^{1}$ and $\vec{d}^{2}$ constants, i.e.,
\begin{equation}
\begin{split}
&Q(M^{1},M^{2})\\
=&\frac{P(A^\mathcal{O},\Gamma,M^{1},M^{2}|\vec{d}^{1},\vec{d}^{2})}{\sum\limits_{M^{1},M^{2}}P(A^\mathcal{O},\Gamma,M^{1},M^{2}|\vec{d}^{1},\vec{d}^{2})}\\
=&\frac{\mathbbm{1}_{\{\varphi(M^{1},M^{2})=A^\mathcal{O}\}}\cdot\mathbbm{1}_{\{\Gamma^1\in M^{1}, \Gamma^2\in M^{2}\}}\cdot\prod\limits_{i<j}\prod\limits_{\alpha=1}^2[\frac{\vec{d}^\alpha(i)\cdot\vec{d}^\alpha(j)}{||\vec{d}^\alpha||_1-1}]^{M^{\alpha}_{ij}}\cdot[1-\frac{\vec{d}^\alpha(i)\cdot\vec{d}^\alpha(j)}{||\vec{d}^\alpha||_1-1}]^{1-M^{\alpha}_{ij}}}{\sum\limits_{M^{1},M^{2}}\mathbbm{1}_{\{\varphi(M^{1},M^{2})=A^\mathcal{O}\}}\cdot\mathbbm{1}_{\{\Gamma^1\in M^{1}, \Gamma^2\in M^{2}\}}\cdot\prod\limits_{i<j}\prod\limits_{\alpha=1}^2[\frac{\vec{d}^\alpha(i)\cdot\vec{d}^\alpha(j)}{||\vec{d}^\alpha||_1-1}]^{M^{\alpha}_{ij}}\cdot[1-\frac{\vec{d}^\alpha(i)\cdot\vec{d}^\alpha(j)}{||\vec{d}^\alpha||_1-1}]^{1-M^{\alpha}_{ij}}}\\
=&\prod\limits_{i<j}\frac{\mathbbm{1}_{\{\varphi(M_{ij}^{1},M_{ij}^{2})=A_{ij}^\mathcal{O}\}}\cdot\mathbbm{1}_{\{\Gamma^1_{ij}\in M_{ij}^{1}, \Gamma^2_{ij}\in M_{ij}^{2}\}}\cdot\prod\limits_{\alpha=1}^2[\frac{\vec{d}^\alpha(i)\cdot\vec{d}^\alpha(j)}{||\vec{d}^\alpha||_1-1}]^{M^{\alpha}_{ij}}\cdot[1-\frac{\vec{d}^\alpha(i)\cdot\vec{d}^\alpha(j)}{||\vec{d}^\alpha||_1-1}]^{1-M^{\alpha}_{ij}}}{\sum\limits_{M_{ij}^1,M_{ij}^2}\mathbbm{1}_{\{\varphi(M_{ij}^{1},M_{ij}^{2})=A^\mathcal{O}\}}\cdot\mathbbm{1}_{\{\Gamma^1_{ij}\in M_{ij}^{1}, \Gamma^2_{ij}\in M_{ij}^{2}\}}\cdot\prod\limits_{\alpha=1}^2[\frac{\vec{d}^\alpha(i)\cdot\vec{d}^\alpha(j)}{||\vec{d}^\alpha||_1-1}]^{M^{\alpha}_{ij}}\cdot[1-\frac{\vec{d}^\alpha(i)\cdot\vec{d}^\alpha(j)}{||\vec{d}^\alpha||_1-1}]^{1-M^{\alpha}_{ij}}}.
\end{split}
\end{equation}

In the M-step, we differentiate Eq. (\ref{eq.lb}) with respect to $\vec{d^1}$ and $\vec{d^2}$ while fixing $Q(\mathbf{M})$ as constants, and find the solution to the equations
\begin{equation}
\begin{split}
\frac{\partial}{\partial\vec{d}^{1},\vec{d}^{2}} \sum_{M^{1},M^{2}}Q(M^{1},M^{2})\ln\frac{P(A^\mathcal{O},\Gamma,M^{1},M^{2}|\vec{d}^{1},\vec{d}^{2})}{Q(M^{1},M^{2})}&=0,
\end{split}
\end{equation}
i.e.,
\begin{equation}
\begin{split}
\left\{
             \begin{array}{lr}
             \sum\limits_{M^{1},M^{2}}Q(M^{1},M^{2})\frac{\partial}{\partial\vec{d}^{1}}\ln P(A^\mathcal{O},\Gamma,M^{1},M^{2}|\vec{d}^{1},\vec{d}^{2})=0\\
             \sum\limits_{M^{1},M^{2}}Q(M^{1},M^{2})\frac{\partial}{\partial\vec{d}^{2}}\ln P(A^\mathcal{O},\Gamma,M^{1},M^{2}|\vec{d}^{1},\vec{d}^{2})=0\\
             \end{array}
\right..
\end{split}\label{si.deri}
\end{equation}

Notice that the probability 
\begin{equation}
\begin{split}
P(A^\mathcal{O},\Gamma,M^{1},M^{2}|\vec{d}^{1},\vec{d}^{2})=\mathbbm{1}_{\{\varphi(M^{1},M^{2})=A^\mathcal{O}\}}\cdot\mathbbm{1}_{\{\Gamma^1\in M^{1}, \Gamma^2\in M^{2}\}}\cdot P(M^{1},M^{2}|\vec{d}^{1},\vec{d}^{2}),
\end{split}\label{si.likeli}
\end{equation}
and
\begin{equation}
\begin{split}
P(M^{1},M^{2}|\vec{d}^{1},\vec{d}^{2})=P(M^{1}|\vec{d}^{1})\cdot P(M^{2}|\vec{d}^{2}).
\end{split}\label{si.inde}
\end{equation}
Combing the Eqs. (\ref{si.likeli}), (\ref{si.inde}), and substituting into Eqs. (\ref{si.deri}), we obtain 
\begin{equation}
\begin{split}
\left\{
             \begin{array}{lr}
             \sum\limits_{M^{1},M^{2}}Q(M^{1},M^{2})\frac{\partial}{\partial\vec{d}^{1}}\ln [\frac{\vec{d}^1(i)\cdot\vec{d}^1(j)}{||\vec{d}^1||_1-1}]^{M^{1}_{ij}}\cdot[1-\frac{\vec{d}^1(i)\cdot\vec{d}^1(j)}{||\vec{d}^1||_1-1}]^{1-M^{1}_{ij}}=0\\
             \sum\limits_{M^{1},M^{2}}Q(M^{1},M^{2})\frac{\partial}{\partial\vec{d}^{2}}\ln [\frac{\vec{d}^2(i)\cdot\vec{d}^2(j)}{||\vec{d}^2||_1-1}]^{M^{2}_{ij}}\cdot[1-\frac{\vec{d}^2(i)\cdot\vec{d}^2(j)}{||\vec{d}^2||_1-1}]^{1-M^{2}_{ij}}=0\\
             \end{array}
\right..
\end{split}\label{si.derilast}
\end{equation}
For large networks, we regard the term $||\vec{d}||_1$ as a constant, and we can obtain the solution to the Eqs. (\ref{si.deri})
\begin{equation}
\begin{split}
\vec{d}^\alpha(i)=\sum_{j=1}^NM^\alpha_{i,j},\ \forall i=1,2,\cdots N,\alpha=1,2.
\end{split}\label{si.solution}
\end{equation}

\subsection*{C$\ \ \ $ Estimation theory}\label{si.estimation}
Specifically, supposing that $\theta$ is a parameter (scalar) to be estimated from random variable $x$ submitted to the probability density function $f(x;\theta)$, the variance of any unbiased estimator $\hat{\theta}$ is bounded by the inverse of the Fisher information $I(\theta)$, which is defined by
\begin{equation}
\begin{split}
I(\theta)=\mathbf{E}\left[\left(\frac{\partial \ln f(x;\theta)}{\partial \theta}\right)^2\right]=-\mathbf{E}\left[\frac{\partial^2\ln f(x;\theta)}{\partial\theta^2} \right]
\end{split}.
\end{equation} 

Fortunately, maximum likelihood estimator performs the asymptotic normality, indicating the maximum likelihood estimate converges to a normal distribution when the sample size $N$ approaches the infinity~\cite{newey1994large}, i.e., 
\begin{equation}
\begin{split}
\sqrt N(\hat{\theta}-\theta_0)\rightarrow G(0,I^{-1}(\theta_0)),
\end{split}
\end{equation}
where $\theta_0$ is the real value of parameter.
In the proposed framework, the Fisher information matrix $\mathbf{I}(\mathbf{\Theta})$ is defined as the expectation of Hessian matrix of the logarithmic probability density function, i.e.,
\begin{equation}
\begin{split}
\mathbf{I}(\mathbf{\Theta})_{i,j}=\mathbf{E}_{X}\left[\frac{\partial \ln f(x;\mathbf{\Theta})}{\partial\theta_i}\cdot\frac{\partial \ln f(x;\mathbf{\Theta})}{\partial\theta_j}\right]=-\mathbf{E}_{X}\left[\frac{\partial^2\ln f(x;\mathbf{\Theta})}{\partial\theta_i\partial\theta_j}\right],
\end{split}
\end{equation}
where $x$ is any observation including the aggregate network $A^{\mathcal{O}}$ and partial observations $\Gamma$. 
According to Cramer-Rao inequality, the variance of estimator $\mathbf{D}(\hat{\theta}_i)$ satisfies
\begin{equation}
\begin{split}
\mathbf{D}(\hat{\theta}_i)\geq (\mathbf{I}^{-1}(\mathbf{\Theta}))_{i,i}
\end{split}.
\end{equation}
In other words, the $i$-th element of leading diagonal in matrix $\mathbf{I}^{-1}$ shows the lower bound of the variance of unbiased estimator $\hat{\theta}_i$.
To illustrate the asymptotic behavior of variance of the estimator, we present simulations results are shown in Fig. S3, showing the variance of the proposed estimator reaches the corresponding CRLB asymptoticly.

\subsection*{D$\ \ \ $ Percolation process on interdependent networks}\label{si.percolation.}
We analyze the property of robustness between the real network and the reconstructed network in this note. 
Recent work~\cite{buldyrev2010catastrophic} had made pioneering contribution to calculate the size of the giant mutual connected component $\mu_\infty$ with occupied probability $p$.
We will follow the notations of this paper, indicating a two-layer multiplex network is composed by two networks $A$ and $B$. 
We notice that $G_{A0}(z)=\sum_{k}P_A(k)z^k$, which is the generating function of
the degree distribution of network A, and $G_{A1}(z)=G'_{A0}(z)/G'_{A0}(1)$.
Once a fraction, $1-p$, of randomly chosen nodes are removed, the degree distribution of the remaining nodes is changed~\cite{newman2002spread}.
Theoretically, $\mu_\infty=xg_B(x)=yg_A(y)$, and
\begin{equation}
\begin{cases}
x=g_A(y)\cdot p\\
y=g_B(y)\cdot p
\end{cases}, 
\end{equation}
where 
\begin{equation}
\begin{split}
g_A(p)=1-G_{A0}[1-p(1-f_A)]
\end{split}.
\end{equation}
In the equation above, $f_A$ is a function of $p$ that satisfies the transcendental equation $f_A=G_{A1}[1-p(1-f_A)]$.

Since the above process is taken place on unweighted multiplex networks, we will binary the reconstructed network by the probability distribution $Q(\mathbf M)$. 
For each individual link between nodes $i$ and $j$ in layer $\alpha$, the value $Q_{ij}^\alpha$ describes the probability that there exists a link by the observations.
Thus, we yield an unweighted multiplex network by generating each link submitted to a Bernoulli trial for probability $Q_{ij}^\alpha$.
In addition, we consider the networks that all nodes belong to GMCC, which means $\mu_\infty(1)=1$.
After generating the network by Bernoulli trials for each link, we add links by the rank of link reliability until all nodes are in the giant mutual connected component.
We consider the function $\mu_\infty(p)$ to analyze the robustness of the real multiplex network and reconstructed networks.

\subsection*{E$\ \ \ $ Random walk process in interconnected multiplex networks}\label{si.randomwalk}

We will analyze the property of navigability between the real network and the reconstructed network in this note. 
Recent work~\cite{de2014navigability} had made pioneering contribution to analyze the random walk process in an interconnected multiplex network.
A walker walks along a general network by four types of parameters, which are i) $\mathcal{P}^{\alpha\alpha}_{ii}$, the probability for staying at the same node $i$ and in the same layer $\alpha$; ii) $\mathcal{P}^{\alpha\alpha}_{ij}$, the probability for walking from node $i$ to its neighbor node $j$ in the same layer $\alpha$; iii) $\mathcal{P}^{\alpha\beta}_{ii}$, the probability for switching from layer $\alpha$ to layer $\beta$ while staying at the same node $i$; iv) $\mathcal{P}^{\alpha\beta}_{ij}$, the probability for walking from node $i$ in layer $\alpha$ to node $j$ in layer $\beta$. 
For an interconnected multiplex network such as a multiplex transportation network, $\mathcal{P}^{\alpha\alpha}_{ii}$ indicates the probability that a person stays put without going anywhere, and $\mathcal{P}^{\alpha\beta}_{ii}$ indicates that a person switches the means of transportation but still stays at the same region.
Similarly, $\mathcal{P}^{\alpha\alpha}_{ij}$ indicates the probability that a person goes to node $j$ from node $i$ without changing means of transportation, while $\mathcal{P}^{\alpha\beta}_{ij}$ equals to $0$.

Thus, the probability for finding the
walker at node $j$ in layer $\beta$ at time $t+\Delta t$ is given by
\begin{equation}
\begin{split}
p_{j\beta}(t+\Delta t)=\mathcal{P}^{\alpha\alpha}_{jj}\cdot p_{j\beta}(t)+\sum_{\alpha\neq\beta}\mathcal{P}^{\alpha\beta}_{jj}\cdot p_{j\alpha}(t)+\sum_{i\neq j}\mathcal{P}^{\beta\beta}_{ij}\cdot p_{i\alpha}(t)
\end{split}.\label{eq.randomw}
\end{equation}
Practically, we introduce a real number $p_{inter},0<p_{inter}<1$ to indicate the probability for a person to change vehicle.
Thus, given a multiplex network $\mathbf{M}$, we have
\begin{equation}
\begin{split}
\mathcal{P}^{\alpha\beta}_{ii}=\frac{p_{inter}}{L-1}\ (\forall i, \forall \alpha\neq\beta)
\end{split},
\end{equation}
\begin{equation}
\begin{split}
\mathcal{P}^{\alpha\alpha}_{ij}=\begin{cases}
(1-p_{inter})\cdot\frac{M^\alpha_{ij}}{k^\alpha_{i}},\text{if}\ \ k^\alpha_{i}\neq0\\
0,\text{if}\ \ k^\alpha_{i}=0
\end{cases}(\forall\alpha,\forall i\neq j),
\end{split}
\end{equation}
and
\begin{equation}
\begin{split}
\mathcal{P}^{\alpha\alpha}_{ii}=\begin{cases}
1-p_{inter}, \text{if}\ \ k^\alpha_{i}=0\\
0, \text{if}\ \ k^\alpha_{i}\neq0
\end{cases} (\forall \alpha,\forall i).
\end{split}
\end{equation}
The process described in Eq. (\ref{eq.randomw}) is a Markov process, since
\begin{equation}
\begin{split}
\sum_\beta\sum_j\mathcal{P}^{\alpha\beta}_{ij}=1, \forall \alpha,i
\end{split},
\end{equation}
We mainly study the navigability of the interconnected multiplex network by the coverage $\phi(t)$, which is the regions been visited until time $t$ by the walkers from a random chosen node. 
Note that the region been visited indicates that the node is visited despite of layers. 
For example, the region $i$ is visited if node $i$ in any layer is visited, because the node $i$ in each layer indicates the same region.

\subsection*{F$\ \ \ $ Spreading process in temporal networks}\label{si.spreading}
We first introduce epidemic spreading process in a single-layer network. 
There are many models describing epidemic spreading process, such as the susceptible-infected (SI) model, the susceptible-infected-susceptible (SIS) model and the susceptible-infected-recovered (SIR) model~\cite{bailey1975mathematical}. 
Here we take the SI model as an illustration.

In a temporal network, the topology may change at each time $t$. 
Thus, we employ a multiplex network $\mathbf{M}$ to describe the time-varying topology, where layer $M^\alpha\ (\alpha=1,2,\cdots T)$ indicates the adjacency matrix at time $t=\alpha$.
Then, we study the spreading of an epidemic disease with SI model in such a multiplex network, indicating each node $i$ of the network has only two states, ``susceptible'' or ``infected'' ($z_i=S\ \text {or}\ z_i=I$). 

A susceptible node is a temporarily healthy node, which can be infected by any infected neighbor node.
Once node $i$ is infected at time $t=\beta$, its state changes to ``susceptible'' since time $t=\beta$. 
At each step, an infected node will infect the susceptible nodes connected to it with probability $\lambda$, which is called the infectious rate. 
Since link reliability $Q^\alpha_{ij}$ describes the probability that there is a link from node $i$ to $j$, the probability for node $j$ being infected by node $i$ at time $t=\alpha$ will be $\mathbf{1}_{\{z_i=I\}}\cdot \lambda\cdot Q^\alpha_{ij}$. 
Then, the probability for a susceptible node $j$ being infected at time $t=\alpha$ is $1-\prod_i(1-\mathbf{1}_{\{z_i=I\}}\cdot\lambda\cdot Q^\alpha_{ij})$.
We consider the proportion of infected nodes $I(T)$ at time $t=T$, i.e., $I(T)={\sum_i\mathbf{1}_{\{z_i=I\}}}/{N}$ to study the spreading process in such multiplex networks.


\clearpage
\section*{References}

\appendix

\beginsupplement

\bibliography{cited}

\end{document}